\documentclass[11pt]{article}
\usepackage[utf8]{inputenc}
\usepackage[T1]{fontenc}  % provides \DJ and other glyphs
\usepackage{graphicx,pstricks,pst-node,psfrag,amsthm,amssymb,amsmath,dsfont,bbm,url}
\usepackage[round,compress]{natbib} %for bibtex
\usepackage[a4paper]{geometry}
\usepackage{setspace}
\usepackage{float} %for begin{figure}[H]
\usepackage{subcaption} %subfigure
\usepackage[labelfont=bf]{caption}
\usepackage{authblk}
\usepackage{tikz}
\usetikzlibrary{positioning}
\usepackage{multirow}
\usepackage{makecell}
\usepackage{adjustbox}

\usepackage{xcolor}
\newcommand{\review}[1]{\textcolor{black}{#1}}

\usepackage{graphicx}
\usepackage{float}
\usepackage{array}
\newcolumntype{C}[1]{>{\centering\arraybackslash}m{#1}}

\include{macros}

\allowdisplaybreaks[1]

\begin{document}

%\title{The hidden cost of stringent motion scrubbing}
\title{Bayesian brain mapping: a population-informed framework for personalized functional network topography and connectivity}
%\author{Nohelia Da Silva Sanchez, Diego Derman, Damon D. Pham, Ellyn R. Butler, Mary Beth Nebel, Amanda F. Mejia} 
\date{}

\author[1]{Diego Derman, PhD*}
\author[1]{Nohelia Da Silva Sanchez*}
\author[1]{Saige Rutherford, PhD}
\author[1]{Damon D. Pham}
\author[2]{Ellyn R. Butler, MS}
\author[3]{Mary Beth Nebel}
\author[1]{Amanda F. Mejia, PhD}

\affil[1]{Department of Statistics, Indiana University}
\affil[2]{Department of Psychology, Northwestern University}
\affil[3]{Center for Neurodevelopmental and Imaging Research, Kennedy Krieger Institute}

%TC:ignore
\date{}
\maketitle

\noindent *Contributed equally

\vspace{1cm}

\noindent Co-author e-mail addresses: ellynbutler2027@u.northwestern.edu, nebel@kennedykrieger.org, afmejia@iu.edu, dderman@iu.edu, ndasilv@iu.edu

\vspace{1cm}

\noindent ORCIDs: E.R.B. (0000-0001-6316-6444), A.F.M. (0000-0002-4312-8974), D.D. (0000-0002-6895-9248)

\vspace{1cm}

\noindent Corresponding author: Diego Derman 

\noindent E-mail: dderman@iu.edu

\doublespacing

\newpage 

\begin{abstract}

The spatial topography of functional brain organization is increasingly recognized to play an important role in cognition and disease. Accounting for individual differences in functional topography is also crucial for accurately distinguishing spatial and temporal aspects of functional brain connectivity. Yet, accurate estimation of personalized functional brain networks from functional magnetic resonance imaging (fMRI) without extensive scanning remains challenging due to high noise levels. Here, we describe Bayesian Brain Mapping (BBM), a technique for personalized functional topography and connectivity informed by population information. BBM relies on population-derived priors on both spatial topography of networks and between-network functional connectivity to guide subject-level estimation and combat noise. These priors are based on existing spatial templates, such as parcellations or continuous network maps, providing correspondence to those templates. Yet BBM is highly flexible, avoiding strong spatial or temporal constraints and allowing for overlap between networks and heterogeneous patterns of engagement. BBM is designed for single-subject analysis, making it computationally efficient and translatable to clinical settings. Here, we describe the BBM model and illustrate the use of the BayesBrainMap R package to construct population-derived priors, fit the model, and perform inference to identify engagements. A demo is provided in an accompanying Github repo. We also share priors derived from the Human Connectome Project and provide code to support the construction of priors from different data sources, lowering the barrier to adoption of BBM for studies of individual brain organization.

\end{abstract}

\newpage

\section{Introduction}

The functional organization of the brain is highly individualized, both in terms of the spatial configuration \citep{gordon2017individual, braga2017parallel, cui2020individual} and temporal dynamics \citep{finn2015functional} of functional brain networks. Individual features of functional network topography are predictive of cognition, disease severity, psychopathology, socio-economic status, and neurodevelopment \citep{bijsterbosch2018relationship, kong2019spatial, bijsterbosch2019relationship, cui2022linking, li2022atlas, keller2023personalized, lynch2024frontostriatal, butler2025sex, pang2026personalized, herzberg2024elabeses, zhao2024sestopography}, \review{and are trait-like, heritable, influenced by both environmental and genetic factors, and composed of multiple phenotypically relevant types of variants \citep{seitzman2019trait, dworetsky2024two}. 
Functional network topography is thus a valuable marker of macro-scale neurological diversity with potential utility for tracking disease progression, elucidating treatment mechanisms, and characterizing neuropsychiatric conditions.}  

Extracting functional topography accurately at the individual level has historically relied on collecting hours of resting-state functional magnetic resonance imaging (fMRI) from each participant \citep{braga2017parallel, gordon2017precision, xue2021detailed}. These precision neuroimaging studies have been crucial in revealing the existence, nature, and extent of individual differences in functional topography. However, translational goals, including using functional topography and connectivity features as biomarkers and for personalized treatment, depend on extracting these features accurately from moderate amounts of data feasible to collect in clinical settings.

Another motivation for extracting individualized functional topography from standard-duration scans is its relevance to measurement of FC, i.e. the temporal synchrony between regions of the brain. The use of group parcellations or network maps misaligned to the individual's functional topography can lead to biased FC estimates due to mixing of signals from distinct functional areas \citep{smith2011network, bijsterbosch2018relationship, bijsterbosch2019relationship, butler2019depressive}. Thus, true differences in functional topography can be misinterpreted as differences in FC. A potential ramification is that real relationships between behavioral or phenotypic measures and functional topography may be incorrectly attributed to FC. If prediction is the main objective, such mixing of underlying features may not be highly consequential. But if we aim to understand the brain mechanisms behind certain behaviors or disease states, rather than simply predict them, it is crucial to disentangle differences in spatial topography from differences in temporal engagement and connectivity \citep{harrison2020modelling}.

To extract individualized or ``personalized'' functional networks (PFNs) from typical duration subject-level fMRI data, hierarchical Bayesian models have proven an effective approach. These models reduce noise by combining information from multiple subjects, while respecting individual differences and ensuring correspondence between individuals.  Hierarchical models have been successfully applied in the context of parcellation \citep{kong2019spatial, kong2021individual}, probabilistic functional modes (PROFUMO) \citep{harrison2015large, farahibozorg2021hierarchical}, and independent component analysis (ICA) \citep{guo2013hierarchical, mejia2020template}. Importantly, these methods have been shown to perform well using a modest amount of data per individual, potentially avoiding the need to collect prolonged or repeated fMRI sessions in individuals. % \citep{kong2021individual, mejia2020template}.  
This makes them highly attractive in contexts where extensive subject-level scanning is not feasible, including clinical settings where it presents a financial and physical burden for patients.

In addition to the importance of accounting for individual differences in functional topography, there is growing evidence for the existence and relevance of overlapping functional architecture among brain systems or networks \citep{power2013evidence, gratton2018control, faskowitz2020edge, hermosillo2024precision}. This favors ``soft'' parcellations allowing for overlap between functional networks over traditional ``hard'' parcellations \citep{li2018collective, dadi2020fine, bijsterbosch2023evaluating}. Examples of soft parcellation approaches include spatial ICA \citep{beckmann2004probabilistic}, temporal ICA \citep{smith2012temporally}, PROFUMO \citep{harrison2015large}, non-negative matrix factorization \citep{Li2017nonnegative}, gradients \citep{margulies2016situating}, and dictionaries of functional modes \citep{dadi2020fine}. All of these methods use assumptions or constraints to make estimation feasible, though they differ in the specific choice of constraint. For instance, spatial ICA encourages statistical independence between network maps, leading to relatively little overlap between networks. PROFUMO and temporal ICA allow for greater spatial overlap but encourage temporal independence between networks, resulting in low FC between networks \citep{pervaiz2020optimising}. Relaxing these constraints, while continuing to provide sufficient model guidance for estimation accuracy and alignment of features across individuals, may permit both the temporal and spatial features of functional brain organization to be more fully expressed. 

Here, we present Bayesian Brain Mapping (BBM), a pragmatic and flexible hierarchical Bayesian technique for producing PFNs without strong constraints on the spatial or temporal structure of the networks.  BBM begins with a \textit{template}, which can be either a group parcellation or a set of continuous network maps. That template is used as the basis for population-derived priors on spatial topography and FC, which are then used to inform subject-level estimation of those same features.  BBM is a \review{novel} generalization of template ICA \citep{mejia2020template}, which uses group ICA maps as the template, but BBM allows for parcellations, ICA maps, PROFUMO modes, and other types of network maps as the template. \review{Additionally, we introduce the companion \texttt{BayesBrainMap} package, which facilitates broad adoption in applied research settings.}

In this work, we describe BBM, illustrate its use, and provide resources to facilitate its adoption.  We have three primary goals. First, we demonstrate the use of the \texttt{BayesBrainMap} R package, which includes functions for establishing population-derived priors and for fitting subject-level models to estimate individual functional topography and connectivity. Second, we illustrate the flexibility of BBM to work with various choices of template, including parcellations and continuous network maps, and how the priors relax the more constrained templates. Finally, we share high-quality priors derived from the Human Connectome Project (HCP) \citep{van2013wu} using a variety of templates, along with instructions and code to replicate those priors. The priors can be directly adopted to apply BBM to studies of healthy young adults, and our pipeline for generating them can be adapted to produce custom priors based on other templates or populations. A comprehensive demo available on Github accompanies this manuscript and provides further details, examples, and visualizations.

\section{Methods}

\begin{figure}
    \centering
    \includegraphics[width=1\linewidth]{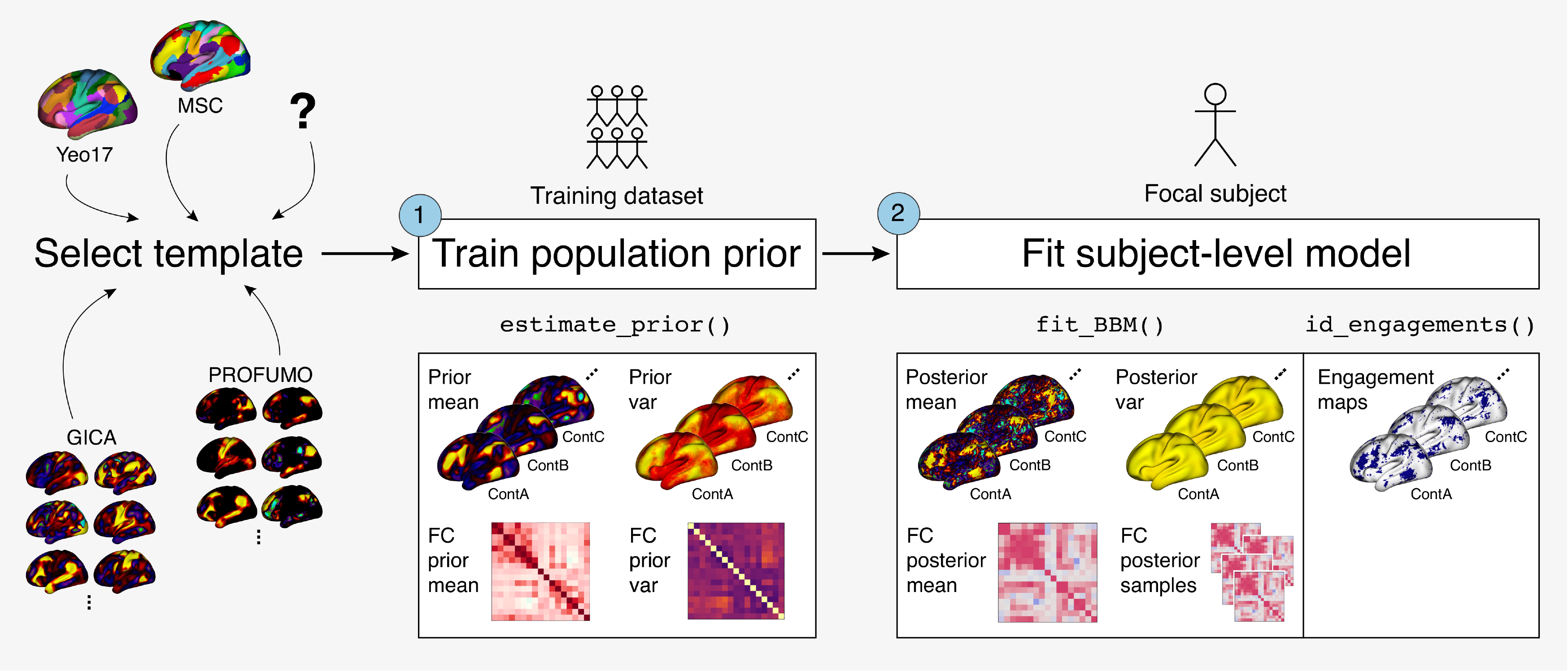}
    \caption{\textbf{Overview of Bayesian Brain Mapping.} The two main steps are (1) prior estimation and (2) model fitting, both implemented in the \texttt{BayesBrainMap} R package. \review{Step 1 is optional since pre-calculated HCP priors are available for download on OSF.}}
    \label{fig:BBM_overview}
\end{figure}

%In BBM, the template is not used directly in the individual-level model, in contrast to dual regression (CITE), group-information guided ICA (CITE), or constrained ICA (CITE).  
Figure \ref{fig:BBM_overview} illustrates the BBM framework. First, the selected template is used to construct population-derived priors for the functional topography of and FC between networks. The network priors encode inherent population variability in the boundaries, configuration, and overlap of networks, effectively relaxing any strong constraints of the template. \review{The training dataset should be drawn from a representative repository (e.g., the HCP for healthy young adults, the Adolescent Brain Cognitive Development study \citep{casey2018adolescent} for adolescents, or the Alzheimer's Disease Neuroimaging Initiative \citep{mueller2005alzheimer} to include Alzheimer's subjects).} %For more specialized populations, the training set can be drawn from the focal study. If insufficient data is available to hold out for training, the data to be analyzed can be used to estimate the prior---an empirical Bayes approach that comes with higher risk of overfitting.
Second, the priors are used to guide and inform estimation in a subject-level Bayesian model to obtain posterior estimates of individual functional topography and FC, as well as measures of uncertainty that can be used for inference. Note that there is no need to fit a large multi-subject hierarchical model. The ability to analyze data from a single subject at a time makes BBM highly pragmatic, computationally efficient, parallelizable, and feasible in clinical settings.

%The BBM model is hierarchical in the sense that it includes population-derived priors on the parameters representing subject-specific spatial topography of different networks and their corresponding temporal activation profiles. However, while hierarchical models typically require multi-subject data to establish the shared prior parameters \citep{guo2013hierarchical, harrison2015large}, in BBM the priors are established first, then the BBM model can be fit to data from a single subject at a time. This allows BBM to be highly pragmatic, computationally efficient, parallelizable, and clinically feasible.

The BBM population-derived priors reduce noise while retaining signal, resulting in more reliable and accurate personalized functional network maps and the FC between them \citep{mejia2020template, mejia2023template, mejia2025leveraging}. An important feature of the functional topography priors is that the population variance of each network varies across the brain, generally showing greater inter-individual differences in regions of high engagement, and greater similarity across subjects in background regions where engagement is low, \review{as observed within high-intensity areas in the prior means and variances (\texttt{estimate\_prior()} panel in Figure \ref{fig:BBM_overview})}. This allows individual differences to be expressed where they exist, while reducing noise in background regions through shrinkage. Due to the continuous, whole-brain nature of BBM network maps, noise reduction in background regions of one network also indirectly contributes to estimation of signal in the same area of the brain in other networks. 

BBM includes an optional population-derived prior on the FC between networks to encode population trends and variability in FC, which shrinks the subject-level FC toward the population mean to alleviate sampling error and stabilize within-subject fluctuations. It is well-established that Bayesian shrinkage is beneficial for reliable FC estimation \citep{ledoit2004well, su2009modified, chen2010shrinkage, varoquaux2010brain, shou2014shrinkage, mejia2015improving, dai2017predicting, mejia2018improved, rahim2019population, pervaiz2020optimising, honnorat2022covariance}. However, the inverse-Wishart distribution, a common choice of parametric prior for FC due to its conjugacy with the multivariate Normal distribution \citep{honnorat2019covariance, harrison2020modelling, mejia2025leveraging}, has limited ability to capture population variance patterns. We recently developed a novel informative prior for correlation matrices based on Cholesky factorization, which accurately captures population variance in FC and outperforms the inverse-Wishart for FC estimation \citep{mejia2025leveraging}.

    In this section, we describe and illustrate the two main steps in BBM: prior estimation and model fitting. We first present the Bayesian model. % for individual-level functional topography and connectivity with population priors. 
We then describe how we derive the priors, illustrating the process using HCP data. Our HCP-derived priors are freely available for download, and we share our code so it can be adapted to other contexts. Finally, we illustrate the application of BBM to several individuals from the HCP. All of the techniques described here are implemented in the \texttt{BayesBrainMap} R package, and code for the analyses presented are available via Github (\url{https://github.com/statMINDlab/BBM-demo}).

\subsection{The BBM hierarchical Bayesian model}

For a given subject, let ${y}_{tv}$ be the preprocessed BOLD fMRI time series at location \review{(vertex or voxel)} $v$ and time point $t$, \review{which has been centered and scaled to induce a common scale. Additional details on scaling are provided in Section \ref{sec:priors}.} The BBM model assumes the BOLD time series can be linearly deconstructed into contributions from a set of networks. The model is similar to ICA and other source separation models, but without strong temporal or spatial constraints. 
\begin{align}
\label{eqn:BBM_model}
{y}_{tv} &= \sum_{q=1}^{Q} a_{tq} s_{qv} + {e}_{tv} 
= \mathbf{a}_t^\top \mathbf{s}_v + {e}_{tv}, 
\quad {e}_{tv} \sim N(0, \tau^2_v) \\
s_{qv} &= s^0_{qv} + \delta_{qv},\quad\delta_{qv} \sim N(0, \sigma^2_{qv}) \label{eq:level2} \\
\mathbf{a}_t &\sim N(\mathbf{0}, \mathbf{G}),\quad\mathbf{G} \sim p(\mathbf{G})
\end{align}

At level (1) of the model, $s_{qv}$ is the spatial engagement of network $q$ at location $v$, and $a_{tq}$ is the temporal activation of network $q$ at time $t$. The vectors $\mathbf{a}_t$ and $\mathbf{s}_v$ combine those values across all $Q$ networks. The scale of the residual white noise ${e}_{tv}$ is allowed to vary spatially.  In ICA and other blind source separation analyses of fMRI data, it is common to include additional components to capture structured noise from head motion and other sources.  In BBM, these are not included in the model but are automatically estimated and removed beforehand, \review{as described in \cite{mejia2020template}. The number of nuisance components is determined automatically using Minka's method \citep{minka2001automatic}. If possible}, we recommend decomposition-based denoising \review{prior to BBM} such as ICA-FIX \citep{griffanti2014ica} or ICA-AROMA \citep{pruim2015ica} to reduce structured noise prior to model fitting, as this has been shown to improve performance across methods \citep{harrison2020modelling}. \review{If advanced denoising techniques have already been applied to the data, it may not be necessary to additionally remove nuisance components via BBM. However, we encourage researchers to assess the benefits in their own data.}

Level (2) of the model incorporates the population-derived prior on the spatial topography in ${s}_{qv}$, where the population mean $s^0_{qv}$ and variance $\sigma^2_{qv}$ are considered known, and the deviation terms $\delta_{qv}$ represent individual differences \citep{mejia2020template}. Spatial dependencies in $\delta_{qv}$ can be modeled via a \review{Gaussian Markov random field} spatial prior for additional accuracy and power, though at a higher computational cost \citep{mejia2023template}. \review{In that case, the prior distribution on $\mathbf{\delta_q}$ is a multivariate Normal distribution, the covariance matrix of which is a combination of the population-derived priors, which encodes the variance, and a unit-variance stochastic partial differential equation (SPDE) spatial process, which encodes the spatial dependence. The second level of the model in equation (\ref{eq:level2}) becomes}

\begin{equation}
\review{\mathbf{s}_{q} = \mathbf{s}^0_{q} + \boldsymbol{\delta}_{q},\quad\mathbf{\boldsymbol\delta}_{q} \sim N(\mathbf{0}, \mathbf{D}_{q}\mathbf{R}_{q}\mathbf{D}_{q}), }\label{eq:stICA} 
\end{equation}

\review{where $\mathbf{R}_q$ is the correlation matrix arising from an SPDE spatial process and $\mathbf{D}_q$ is the diagonal matrix containing the standard deviations ${\sigma_{qv}}$ from the population-derived prior, analogous to the first model in equation (\ref{eq:level2}).}

Level (3) of the model incorporates a population-derived prior on the FC between networks and is optional. This is accomplished by assuming a multivariate Normal prior on $\mathbf{a}_t$ with mean zero and \review{$Q\times Q$} covariance \review{matrix} $\mathbf{G}$, and assuming a population-derived hyperprior on $\mathbf{G}$ \citep{mejia2025leveraging}. Note that we constrain each column of the mixing matrix to unit variance (for identifiability, as often required in blind source separation), so $Cov(\mathbf{a}_t)\equiv Cor(\mathbf{a}_t)$. Hence, the covariance of $\mathbf{a}_t$ is equal to the Pearson correlation between the time courses of each network and represents the between-network FC. We provide two choices for the FC prior: the conjugate inverse-Wishart distribution, which is also used in PROFUMO \citep{harrison2020modelling}, or a novel Cholesky prior developed for this model to more accurately encode population variance patterns \citep{mejia2025leveraging}.  The conjugate prior allows for faster computation, while the Cholesky prior improves performance but is more computationally demanding, since it requires sampling from the prior. 

\subsection{Population-derived priors}
\label{sec:priors}

Here, we describe the construction of population-derived priors using training data from the HCP and illustrate the process via the \texttt{BayesBrainMap} R package, \review{using the \texttt{estimate\_prior()} function}.  All steps are detailed in a demo available at \url{https://github.com/statMINDlab/BBM-demo}.  The HCP-derived priors themselves are available through Open Science Framework (OSF) % \citep{daSilva2026_bbm_priors} 
as described in the Github repo README.  These priors can be directly utilized for application of BBM to analyze healthy young adults. %, as described in Section \ref{sec:application}.  
While we do not advise applying HCP-derived priors directly to study individuals from clearly distinct populations, such as children, the elderly, or individuals with known neurological conditions, our workflow can be replicated using the process and scripts provided to produce priors for other populations.

We consider several different choices of template, including parcellations and continuous network maps (Table \ref{tab:template-summary}). For each template, we build priors with and without global signal regression (GSR), since the decision to use GSR remains the subject of debate. As templates, we use the 17-network Yeo parcellation \citep{yeo2011organization}, the Midnight Scan Club (MSC) group parcellation \citep{gordon2017precision}, HCP-derived group ICA maps with 15 to 50 components \citep{smith2013functional}, and group-level PROFUMO modes \citep{harrison2020modelling}. Note that whether the template is a hard parcellation or continuous, the BBM priors and individual functional topography maps are always continuous. % Because the functional MRI data in the HCP was acquired with two different phase-encoding directions (left-to-right, LR, and right-to-left, RL), we build priors using each phase encoding direction separately, as well as a combined version. Comparing the LR and RL priors allows us to assess the impact of phase encoding direction on the priors, while the combined version provides a general purpose prior that is not specific to either acquisition.

\begin{table}
\centering
\begin{tabular}{|l|c|c|c|c|c|c|}
\hline
\textbf{GSR} & \multicolumn{4}{c|}{\textbf{Network Maps}} & \multicolumn{2}{c|}{\textbf{Parcellations}} \\
\cline{2-7}
 & GICA 15 & GICA 25 & GICA 50 & PROFUMO 12 & Yeo 17 & MSC 17 \\
\hline
With GSR & \checkmark & \checkmark & \checkmark & \checkmark & \checkmark & \checkmark\\
\hline
Without GSR & \checkmark & \checkmark & \checkmark & \checkmark & \checkmark & \checkmark \\
\hline
\end{tabular}
\caption{Templates used for construction of HCP-derived priors available through OSF.}
\label{tab:template-summary}
\end{table}

We first select a high-quality, low-motion, balanced training sample from the HCP to ensure reliable and representative priors. Starting from the full HCP sample of $N=1206$ subjects, we apply several filters. First, we exclude subjects with insufficient scan duration ($<$ 10 min) after dropping the first 15 volumes and excluding volumes with excessive head motion, based on a lagged and filtered version of framewise displacement (FD) appropriate for multiband data \citep{power2019distinctions, pham2023less} with a threshold of 0.5 mm. Second, we exclude any remaining related subjects \review{to ensure independent sampling for prior parameter estimation}, % by using the HCP \texttt{Family\_ID} variable 
retaining the first eligible subject per family after motion and scan-duration filtering. \review{Note that while related participants should be excluded from prior estimation, they can still be included in BBM-based analyses as focal subjects, facilitating the investigation of complex research questions such as the heritability of functional network organization or the presence of subclinical or intermediate phenotypes in relatives of diseased individuals.} Finally, we balance sex within each age group by randomly downsampling the overrepresented sex so that males and females are equally represented within each age bin. 

After applying all filters, our final sample contains 348 subjects. Both runs from the first visit %with different phase-encoding directions (left-to-right, LR, and right-to-left, RL) from each subject 
were then used to train the population-derived priors. Note that priors can also be trained using a single run per subject, in which case each run is split down the middle to produce two pseudo runs. \review{Data denoised with ICA-FIX \citep{griffanti2014ica} was used for prior estimation and model fitting; consequently, no additional removal of nuisance components via BBM was performed. BOLD timeseries were centered and scaled by the local signal mean at each location and multiplied by 100, which induces units of percent signal change. As such, the spatial network engagement values represent percent change at each location, given a 1-unit change in temporal activity of the network. The \texttt{estimate\_prior()} function can alternatively scale by the standard deviation of the time signal, which is useful for data that has previously been centered or had the global signal removed. The function is also equipped to perform masking, nuisance regression, volume censoring, high pass filtering, spatial resampling, and GSR. The results presented here correspond to no GSR, but we also generate and share priors with GSR for the same set of templates via OSF.}

We now briefly describe the process of deriving the priors, all steps of which are implemented in \texttt{estimate\_prior()}. For more details, see the Github demo that accompanies this manuscript. \review{For each subject, either two sessions are provided, or a single session is split to obtain two pseudo-sessions.}
For each training subject and \review{(pseudo-)}session, dual regression \citep{beckmann2009group} is used with the chosen template to produce a noisy set of time courses and spatial maps.\footnote{For templates that are parcellations, we use a modified version of dual regression: note that if network maps were binary and non-overlapping, then the first step of dual regression %(which is simply a multiple linear regression relating the fMRI time series to the network maps treating locations as the data observations and networks as the independent variables) 
would produce the mean time series within each parcel as the corresponding time series.  If group parcels are misaligned to an individual, this will result in mixing signals from multiple networks. To allow for some misalignment between the subject-specific networks and the parcels, we simply compute the median rather than the mean within each parcel. The second step of dual regression then proceeds as usual, as a multiple linear regression modeling the fMRI time series at each location as a function of the network time series to produce a map of engagement for each network.} \review{Dual regression involves first regressing the individual-level BOLD fMRI timeseries against the template maps to obtain their associated individual-level network timeseries; second, the same BOLD timeseries are regressed against those network time series to obtain individual-specific network engagement maps. This results in two sets of network maps and time series for each training subject.} These are then used to estimate the parameters for the topography and FC priors in the BBM model, as we now describe.

Estimation of the topography prior in level (2) of the model is straightforward. \review{Let $i$ index training subjects and $j=1,2$ index (pseudo)-sessions.} Consider a single location \review{(vertex or voxel)} $v$ and network $q$, we aim to estimate the population mean $s^0_{qv}$ and the population variance $\sigma^2_{qv}$. Dropping the $q$ and $v$ subscripts momentarily, let \review{$s_{ij}^{\review{DR}}$} be the dual regression estimate of spatial engagement for training subject $i$ at session $j$. The prior mean $s^0$ and variance $\sigma^2$ are estimated based on a simple measurement error model:
\begin{align}
\review{s_{ij}^{DR}}  & = \review{s_i} + e_{ij}, \quad e_{ij}\stackrel{\text{ind}}{\sim} (0, \sigma^2_e)\\
\review{s_i} & \stackrel{\text{ind}}{\sim} (\review{s^0}, \sigma^2),
\end{align}
where \review{$s_i$} represent the true, noise-free spatial engagement, and the residuals $e_{ij}$ are assumed independent of \review{$s_i$}. %The total variance across the $x_{ij}$ consists of both between-subject or signal variance and within-subject or noise variance, i.e. $\sigma^2_x =  \sigma^2 + \sigma^2_e$, while 
\review{Specifically, the prior mean $s^0$ is simply estimated as the average of $s_{ij}^{DR}$ over all subjects and sessions.} The BBM prior variance corresponds to the between-subject variance in the measurement error model, $\sigma^2$, while $\sigma^2_e$ represents noise levels in the dual regression maps. The variance terms in this model are estimated using standard variance decomposition techniques \review{for measurement error models \citep{shou2013quantifying, mejia2020template}}. % or iterative estimation procedures for linear mixed effects models \review{\citep{laird1982}}.
%An unbiased estimate of $\sigma^2$ is given by $\hat\sigma^2_x - \hat\sigma^2_e$, where $\hat\sigma^2_x = \frac{1}{2}\sum_{j=1}^2 Var_i(x_{ij})$ %is the sample variance across the $x_{ij}$ 
%and $\hat\sigma^2_e=\frac{1}{2}Var_i(x_{i2}-x_{i1})$ (see \cite{mejia2020template} for details).  However, this unbiased estimate may produce invalid negative values due to sampling variability, especially for small values of $\sigma^2_z$ common in background regions. In addition, other values may be under-estimated even if they are positive, resulting in an overly informative prior in some areas. Therefore, we instead use a non-negative, upwardly biased estimate of $\sigma^2_z$ equal to the variance of the $\bar{x}_{i} = \frac{1}{J}\sum_{j=1}^J x_{ij}$, the mean over the $J=2$ sessions for each subject.  The upward bias in this variance estimate is equal to $\sigma^2_e/J$, where $J$ is the number of sessions and therefore goes to zero as we include additional sessions and/or as the noise variance $\sigma^2_e$ decreases due to improved data quality or scan duration. %Collecting these estimates across all locations $v$ and networks $q$ results in ...

Estimation of the FC prior in level (3) of the BBM model is based on the $Q\times Q$ Pearson correlation matrices of the dual regression network time series. As described above, two options for the prior are available: the inverse-Wishart, which is conjugate but has limited flexibility to capture population variance patterns, and a novel informative prior for correlation matrices based on Cholesky factorizations, which we recently developed. See \cite{mejia2025leveraging} for details of the estimation procedure for both priors.  % Our previous analyses found that both choices of prior are beneficial, but the novel permuted Cholesky prior outperforms the inverse-Wishart at identifying reliable FC patterns in individuals \citep{mejia2025leveraging}.

A large training sample from a publicly available resource like the HCP is ideal for prior estimation, as long as the focal subject or study to be analyzed comes from a similar population. However, it is sometimes desirable to build a customized prior for the specific population or context of a given study.  In that case, it is possible to use a holdout set which has been done successfully in previous studies \citep{gaddis2022psilocybin, derman2023individual}. This holdout set should be large enough to accurately estimate prior parameters, ideally 100 or more, or at least several dozen.  It is also possible to use the same subjects for prior estimation and subsequent subject-level analysis \citep{butler2025sex}, an approach known as empirical Bayes (EB). While EB comes with a risk of overfitting, it also allows for using all of the available data for training, generally resulting in more accurate priors, which may be considered worth the tradeoff. 

It is vital to visually inspect the prior mean and variance maps to assess the level of noise present.  \review{Figure \ref{fig:samplesize_yeo17} illustrates the effect of training sample size on the stability of the prior mean and standard deviation.  While the prior mean is stable for even very small training samples (i.e., $n=25$), the prior standard deviation is more sensitive to sample size. We recommend that researchers include data from at least $n=50$ to $100$ training participants for prior estimation, and consider that templates with more networks/parcels generally require larger training samples to produce stable prior parameter estimates.} Besides a large training sample, avoiding over-processing of the training data can also reduce noise levels in the priors. Strategies to avoid over-processing include using lenient or data-driven volume censoring, parsimonious nuisance regression/denoising, and high-pass instead of band-pass filtering \citep{pham2023less, bright2017nuisance, tong2019lowpass}.

\subsection{Fitting the single-subject model}
\label{sec:application}

Once the priors have been estimated, the BBM model can be fit to one or more fMRI sessions from a single individual to estimate functional brain topography and FC. Model fitting is implemented in the BayesBrainMap function \texttt{fit\_BBM()} through an iterative procedure based on expectation-maximization \citep{mejia2020template, mejia2023template} or variational Bayes \citep{mejia2025leveraging}, depending on the specific model. This function produces point estimates (i.e., posterior means) of the subject-specific spatial maps and FC matrices, as well as measures of uncertainty (i.e., posterior variances or samples) to facilitate inference.  

Finally, areas of significant engagement for each network are identified using posterior inference with the function \texttt{id\_engagements()}. \review{Several multiplicity correction options are implemented, including false discovery rate (default) and Bonferroni correction.} Because the BBM framework has relatively high power to detect non-zero engagements, we can specify a minimum effect size of interest for inference. Since the spatial maps are in arbitrary units, we draw inspiration from a common method for thresholding group ICA maps, where values over $z$ standard deviations above the mean across space are commonly displayed, while values below this are considered negligible and suppressed. Following this logic, we set a minimum effect size for inference as the prior mean value corresponding to $z$ standard deviations above the mean. The \texttt{id\_engagements()} function allows the user to specify $z$, or to provide multiple values for $z$ to produce a nested set of significant engagements of different strengths (illustrated in Figure \ref{fig:HCP_subject_Yeo17}). 

In the Github demo that accompanies this manuscript, we illustrate the use of the \texttt{fit\_BBM()} and \texttt{id\_engagements()} functions to analyze several individual participants from the HCP.

% \subsection{Application to highly sampled individuals}

% To illustrate the applicability of BBM when using externally-derived priors, we also analyze
% data from the Midnight Scan Club (MSC) \citep{gordon2017precision}, which features ten densely sampled participants. [How many sessions per subject?  How long in total?]  The MSC focuses on a similar population as the HCP, healthy young adults, but the two datasets
% differ in acquisition methods. Since the population-derived priors used in BBM
% encode the distribution of signal, not noise, priors derived from one study can
% be applied to other studies with differing noise properties, as long as the
% populations are similar. We illustrate this through application of
% HCP-derived priors to data from the MSC. Application to the MSC will also
% illustrate the ability of BBM to reveal individual features of functional
% topography from just a single session of data.

\section{Results}

\subsection{Priors relax spatial constraints of templates and allow network overlap}

Prior means on the spatial network topography for several networks from the Yeo17 template are shown in Figure \ref{fig:Yeo17_priors}. As a hard parcellation, the template is constrained to have no overlap between networks. By contrast, the prior reflects the engagement of the network in the training sample, with no constraint to encourage or enforce spatial independence. This additional flexibility allows the template networks to extend beyond their original boundaries, suggesting the presence of regions that engage in multiple networks and/or individual differences in network boundaries.

Arrows indicate areas that differ markedly between the template parcels and the prior mean maps. For instance, in the Default A network, bilateral, roughly symmetric temporal lobe engagement is seen in the prior, whereas the template parcel only includes right-hemisphere temporal lobe engagement. Another noteworthy example is engagement of Broca's area and other areas commonly associated with language in the temporal parietal network, while the corresponding parcel is constrained to the temporal lobe. The prior mean for this network is similar to patterns often seen in networks characterized by strong engagement in the superior temporal gyrus and the temporal parietal junction when using ICA, PROFUMO, or other network estimation methods that allow for overlap between networks.

% ----------------------------------------------------
%Figure 2 -- Yeo17 components and prior means. Show how the prior relaxes the constraints of the template, specifically how we see additional areas of engagement, and more overlap between networks.
% ----------------------------------------------------

%TO DO:
% - Make the network maps the same color -- yellow is hard to see (after preprint), and/or 
% - show white outlines on the prior mean maps (after preprint)

%macros for template and prior mean maps for Fig 2. Change origin directory for Figure2_Yeo17 for different colors for each network.
\newcommand{\PictureA}[1]{
    \includegraphics[width=6cm, trim = 0 12cm 0 3cm, clip]{Figure2_Yeo17/#1}
}
\newcommand{\PictureB}[1]{
    \includegraphics[width=6cm, trim = 0 14.5cm 0 3cm, clip]{Figure2_Yeo17/#1} 
}

% I do not understand why this is needed because the arrow pictures are the exact same as Picture B
\newcommand{\PictureC}[1]{
    \includegraphics[width=6cm, trim = 0 11.5cm 0 1.5cm, clip]{Figure2_Yeo17/#1} 
}
%macro for row labels 
\newcommand{\RotLabel}[1]{
  \begin{picture}(0,60)\put(0,30){\rotatebox[origin=c]{90}{#1}}\end{picture}
}

\begin{figure}
\centering
\small
\begin{tabular}{ccc}
 & \textbf{Template} & \textbf{Prior Mean} \\[10pt]
 \hline\\[-6pt]
{\RotLabel{ContA}} &
{\PictureA{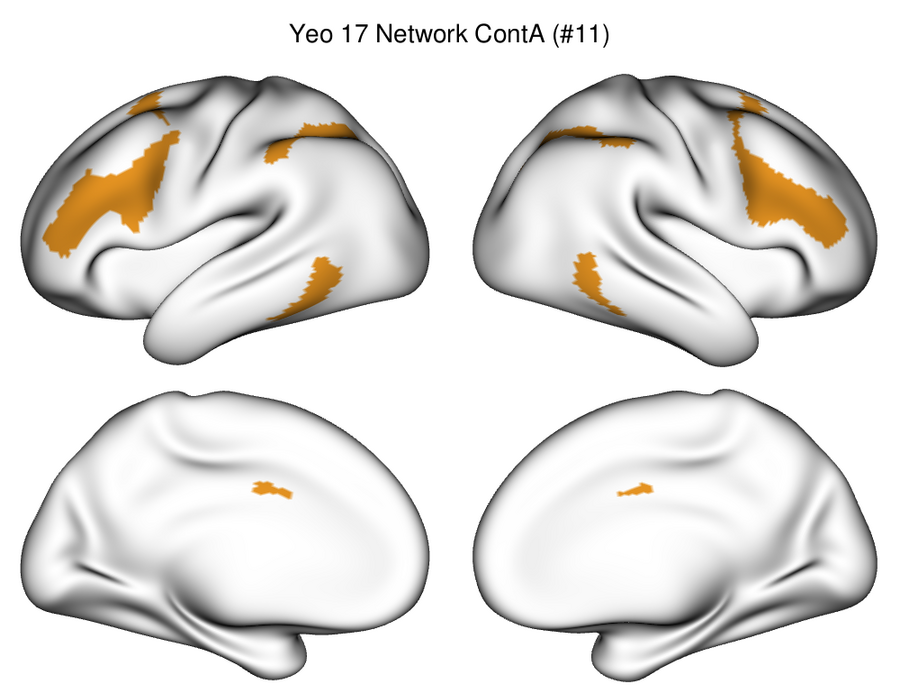}} &
\PictureB{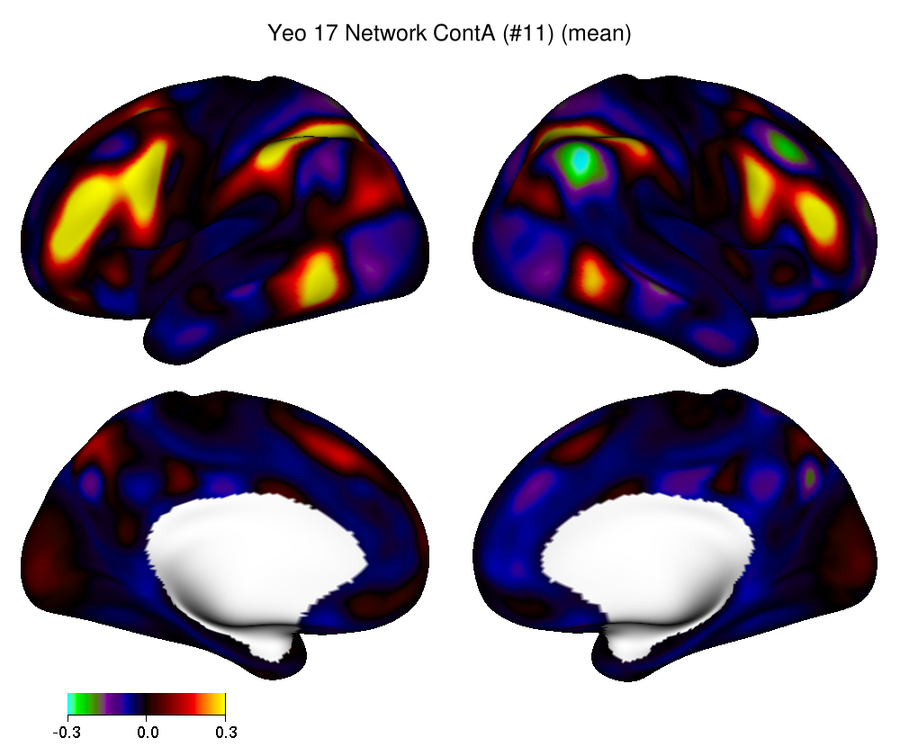} \\
 \hline\\[-6pt]
\RotLabel{DefaultA} &
\PictureA{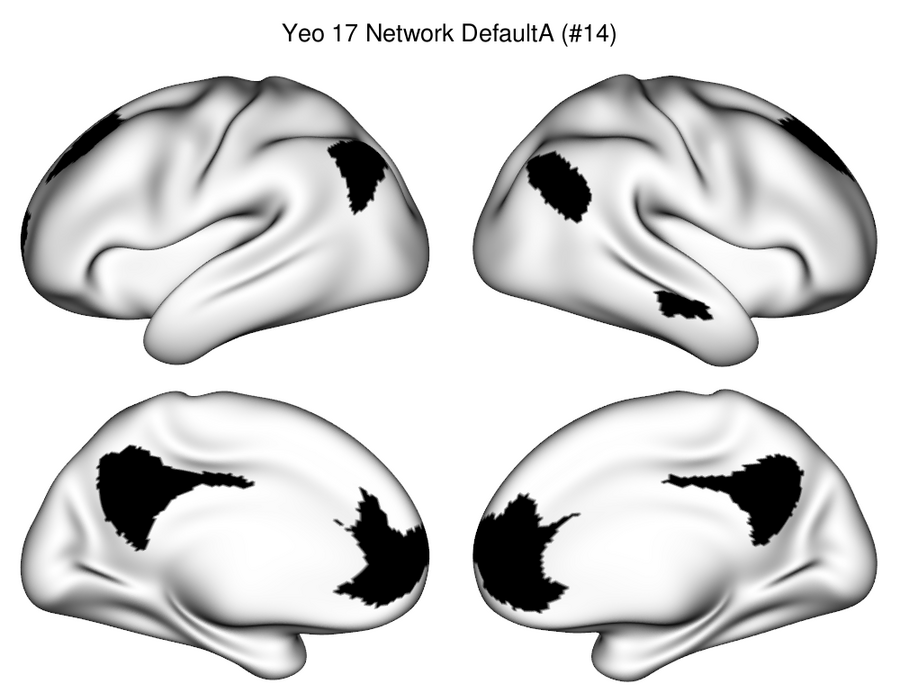} &
\PictureC{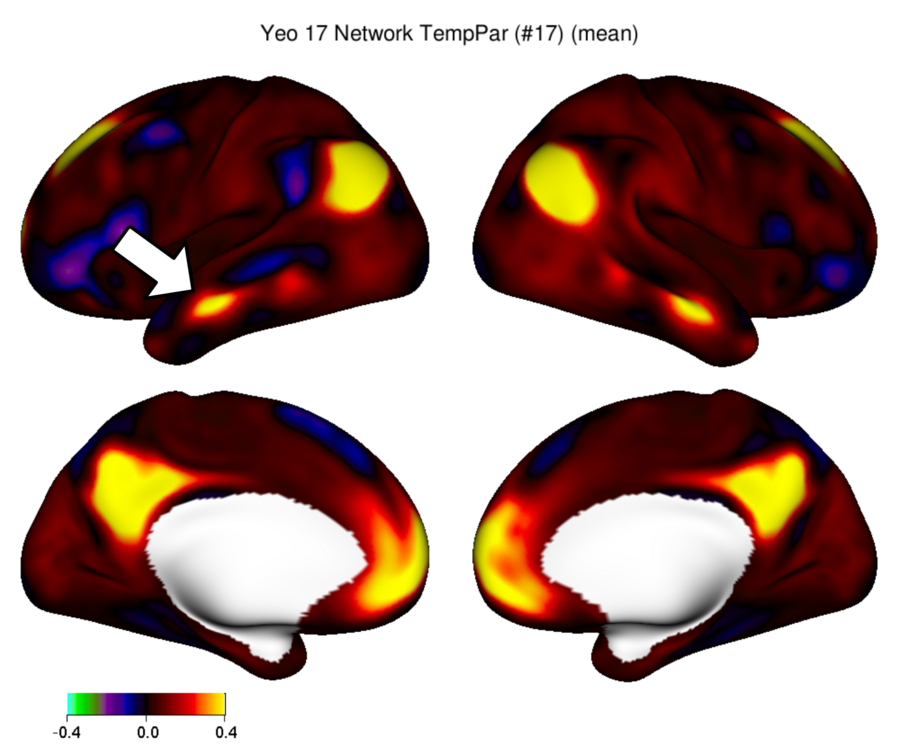} \\
 \hline\\[-6pt]
\RotLabel{DorsAttnA} &
\PictureA{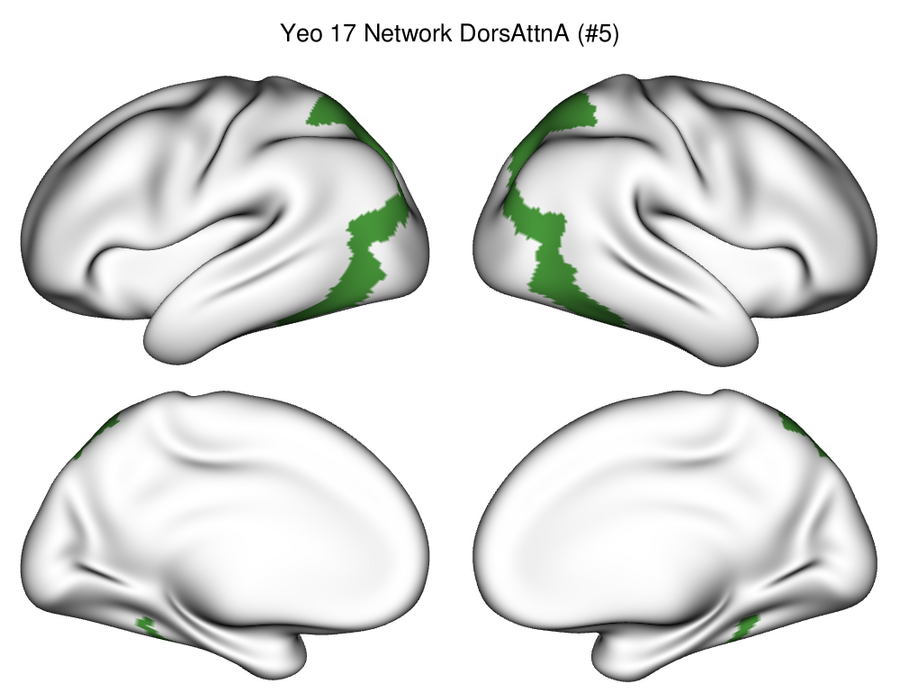} &
\PictureB{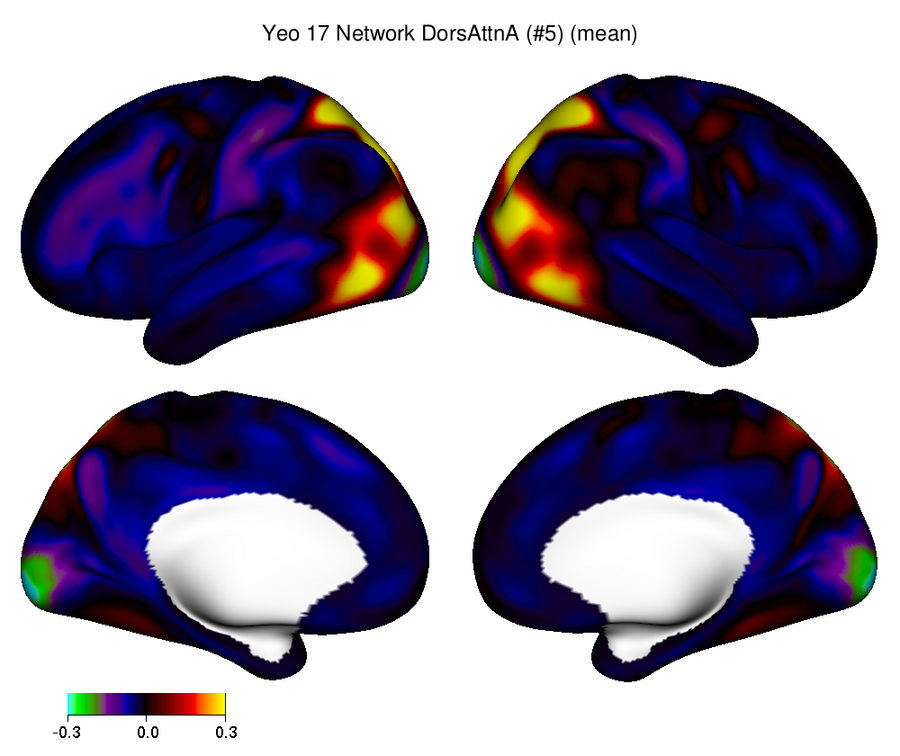} \\
% \hline\\[-6pt]
% \RotLabel{Sal Vent Attn A} &
% \PictureA{Yeo17_SalVentAttnA.png} &
% \PictureB{prior_combined_Yeo17_noGSR_SalVentAttnA_mean.png} \\
\hline\\[-6pt]
\RotLabel{SomMotA} &
\PictureA{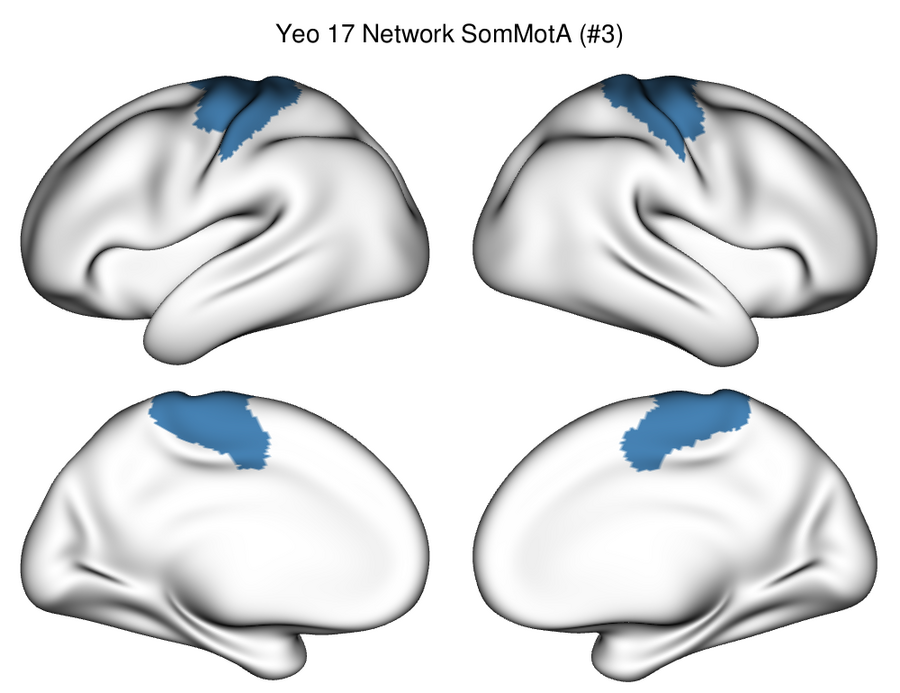} &
\PictureB{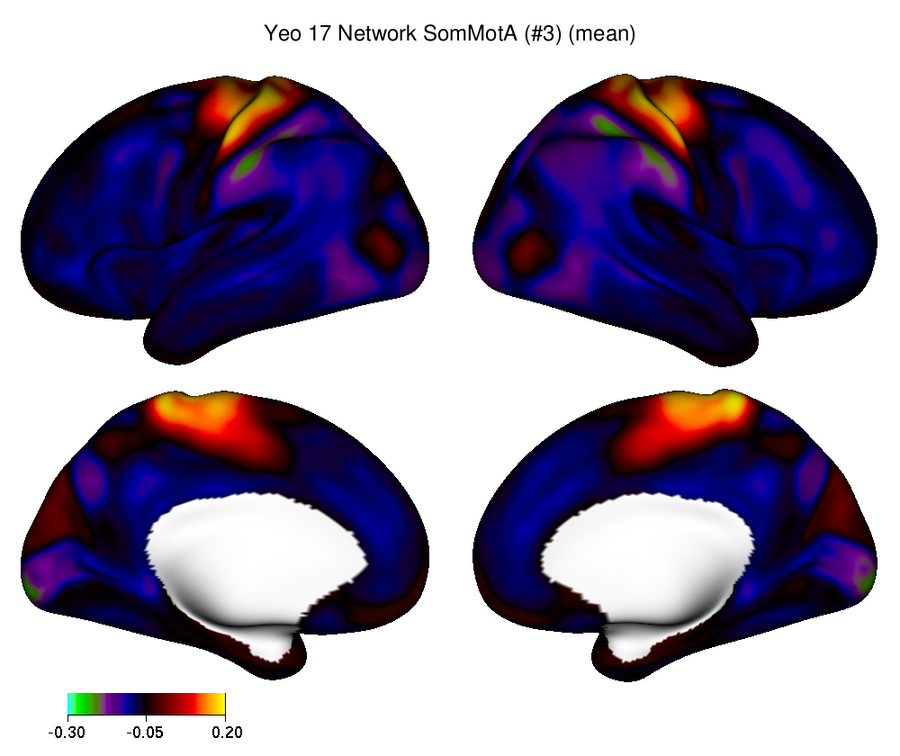} \\
\hline\\[-6pt]
\RotLabel{TempPar} &
\PictureA{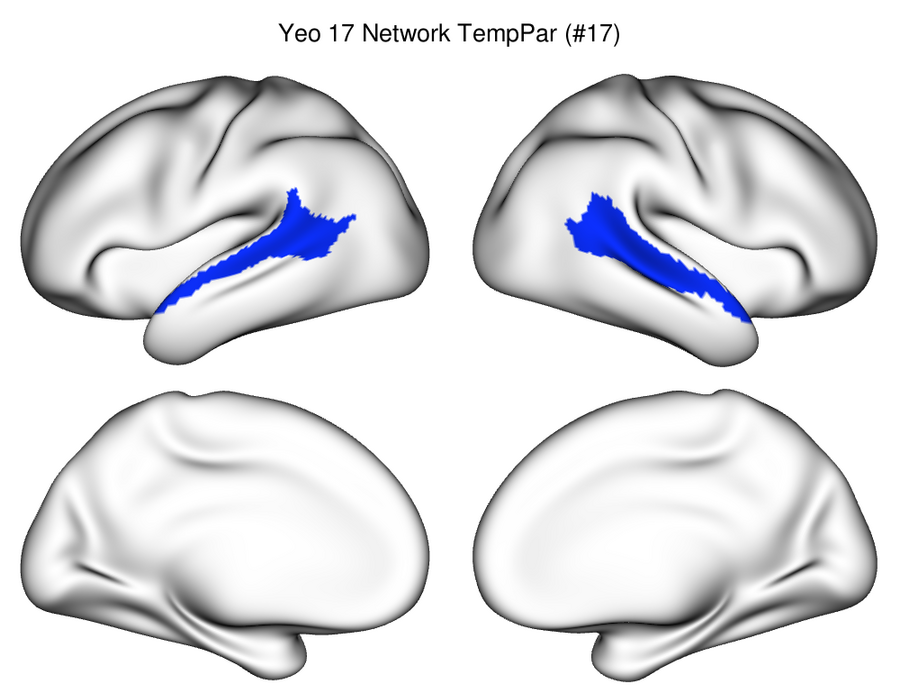} &
\PictureC{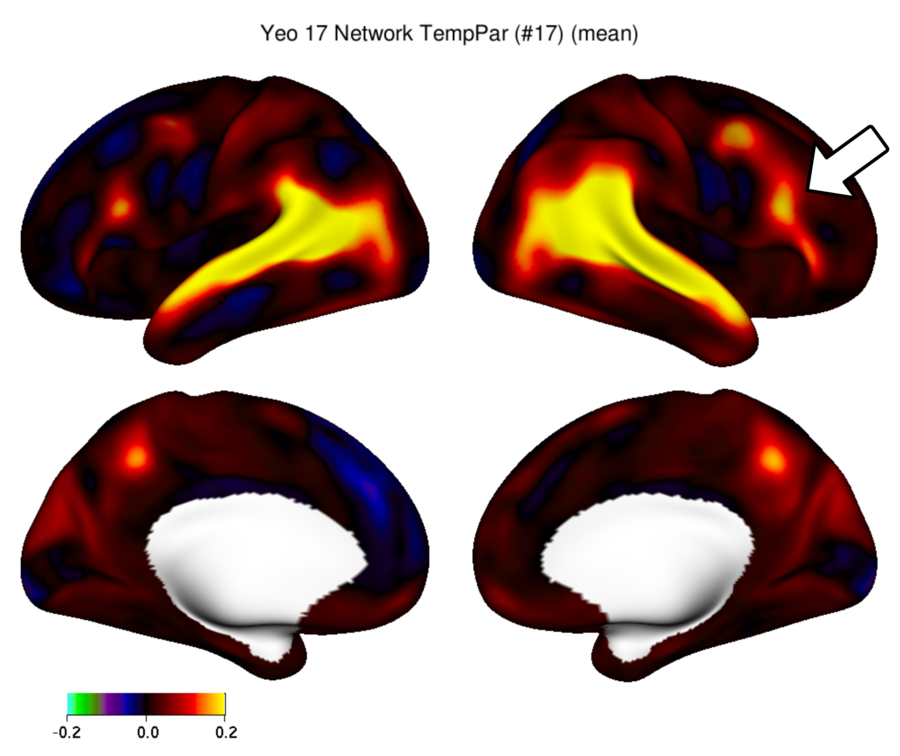} \\
\hline\\[-6pt]
\RotLabel{VisCent} &
\PictureA{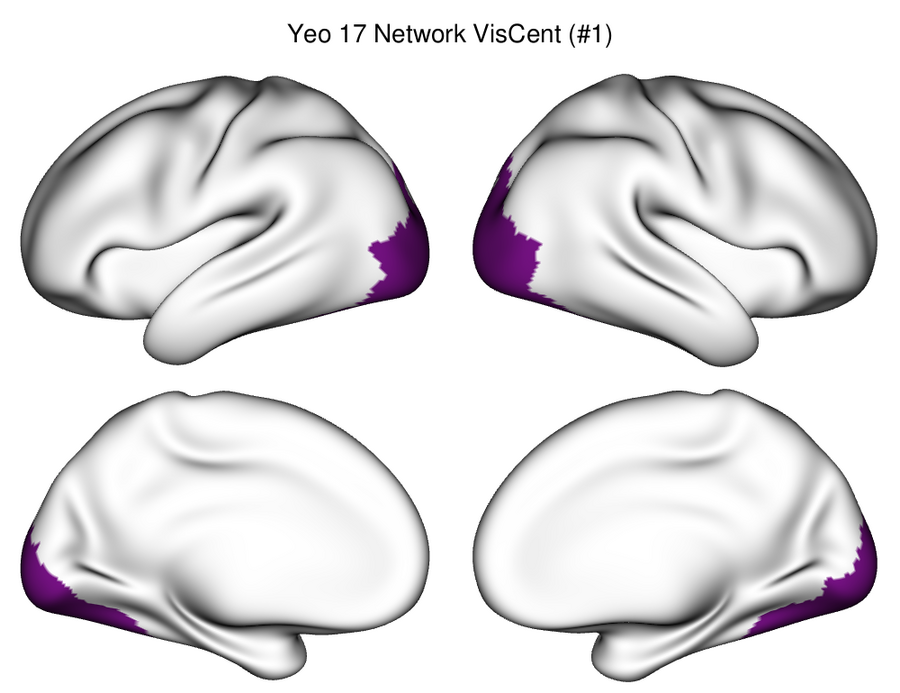} &
\PictureB{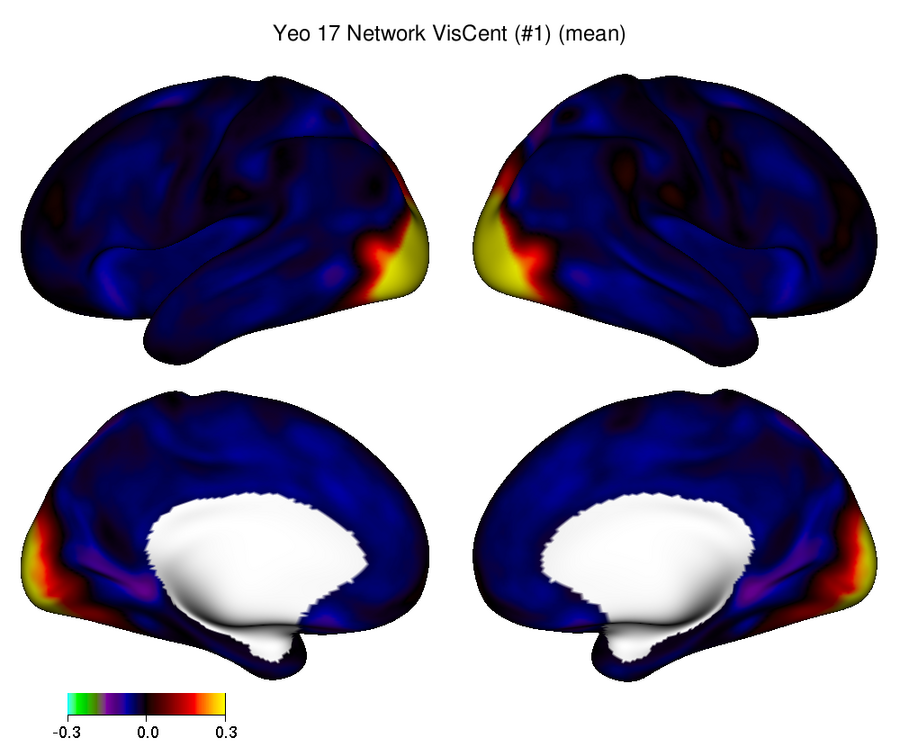} \\
\hline\\[-6pt]
\end{tabular}
\caption{\textbf{HCP-derived spatial topography priors for Yeo17 template.} Six exemplar networks are shown. While the parcels are constrained to have no overlap, the prior mean maps for each network show expansion beyond the parcel boundaries, indicating overlapping network engagement. White arrows denote some areas of significant engagement that overlap different networks and therefore are not present in the original template but can instead be identified by the use of the population-based prior. }
\label{fig:Yeo17_priors}
\end{figure}

% ----------------------------------------------------
%Appendix Fig -- Dice overlap figure.  
% ----------------------------------------------------

%TO DO for figure (after preprint):
% - Use a different color scale that starts at white.  Magma from viridis almost gets to white, but not quite. We could manually add white using color_ramp_palette.  Ideally the chosen color scale would have enough distinct hues to allow us to easily distinguish certain benchmark values, like 50% (after preprint)
% - Show some examples of thresholded maps at |z| > 2 for PROFUMO and GICA (after preprint)

The pairwise spatial overlap between networks for different templates and their corresponding priors is shown in Appendix Figure \ref{fig:spatial_overlap}. Overlap is quantified using the Dice coefficient. For continuous network templates and prior means, we first threshold the maps at a z-score of $\pm 2$ (where the z-score is based on the mean and standard deviation of the prior mean map across all vertices), following a common practice for thresholding group ICA maps to isolate areas of engagement. The parcellation templates, by definition, have zero overlap between networks, yet their corresponding priors exhibit substantial overlap between certain networks. Network maps based on PROFUMO exhibit the most overlap, reflecting a key feature of that method, and the degree of overlap is similar in the template and the prior. To a slightly lesser degree, the GICA template and prior also exhibit overlap. This illustrates that spatial independence in ICA does not prohibit spatial overlap, as is sometimes assumed.\footnote{On the contrary, non-overlap implies anti-correlation, so non-correlated spatial maps will tend \review{to} exhibit overlap. Consider this from a probability perspective: disjoint events are never independent, since the occurrence of one event precludes occurrence of the other.} Still, the GICA prior mean maps exhibit somewhat stronger overlap than the GICA templates, suggesting a relaxation of the spatial constraints in the ICA model. %

% ----------------------------------------------------
%Figure 3 -- Spatial topography prior mean and standard deviation maps for different templates. Shows how the prior mean maps are similar across templates for a given network
% ----------------------------------------------------

\begin{figure}
\centering
\small
\begin{tabular}{C{0.05\textwidth}C{0.25\textwidth}C{0.25\textwidth}C{0.25\textwidth}}
 & \textbf{Template} & \textbf{Prior Mean} & \textbf{Prior SD} \\
\noalign{\vskip 2pt}\hline\noalign{\vskip 6pt}
\rotatebox{90}{\makecell[c]{\textbf{PROFUMO}\\($Q=12$)}} &
\includegraphics[width=\linewidth]{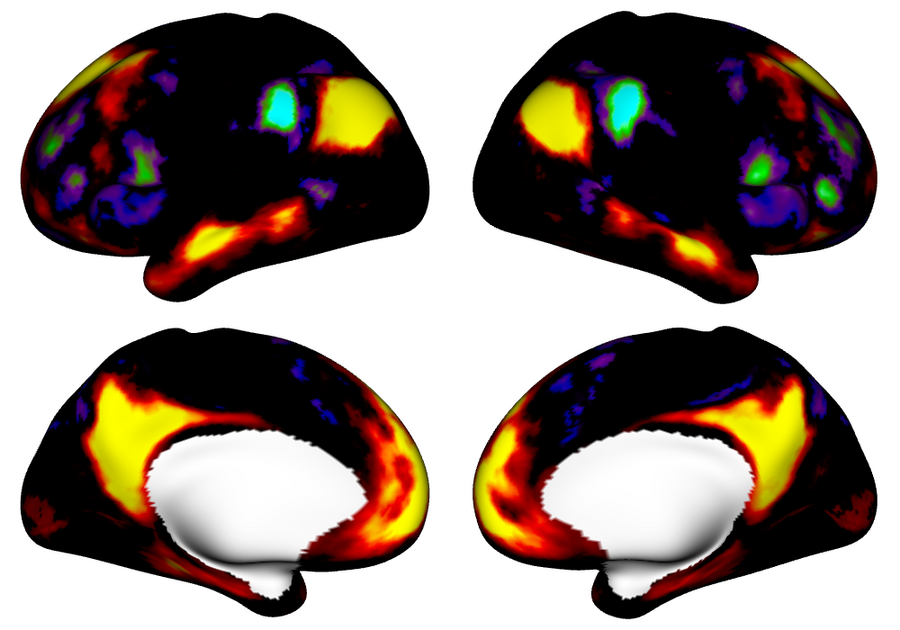} &
\includegraphics[width=\linewidth]{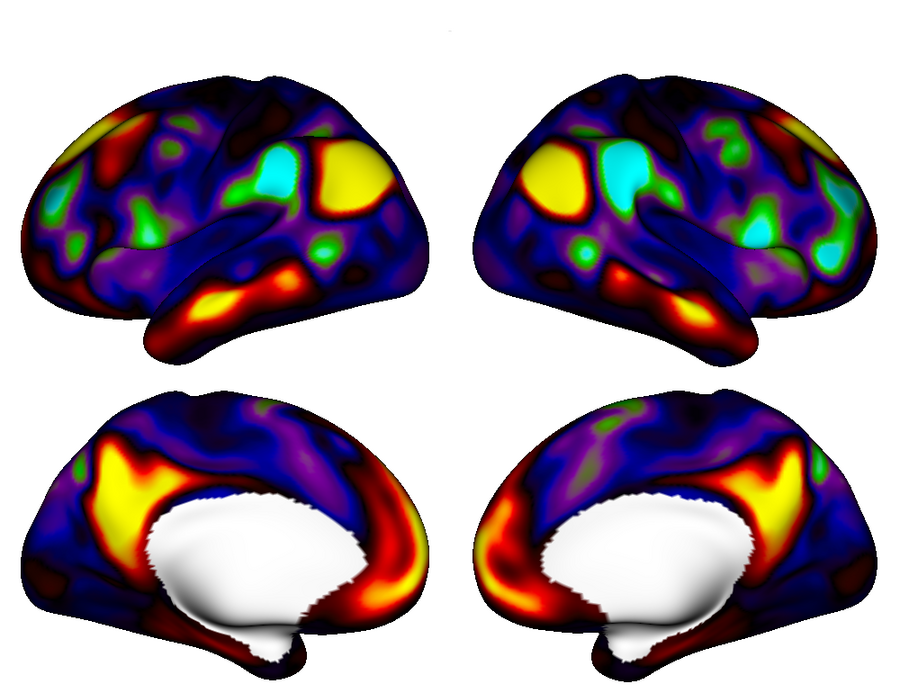} &
\includegraphics[width=\linewidth]{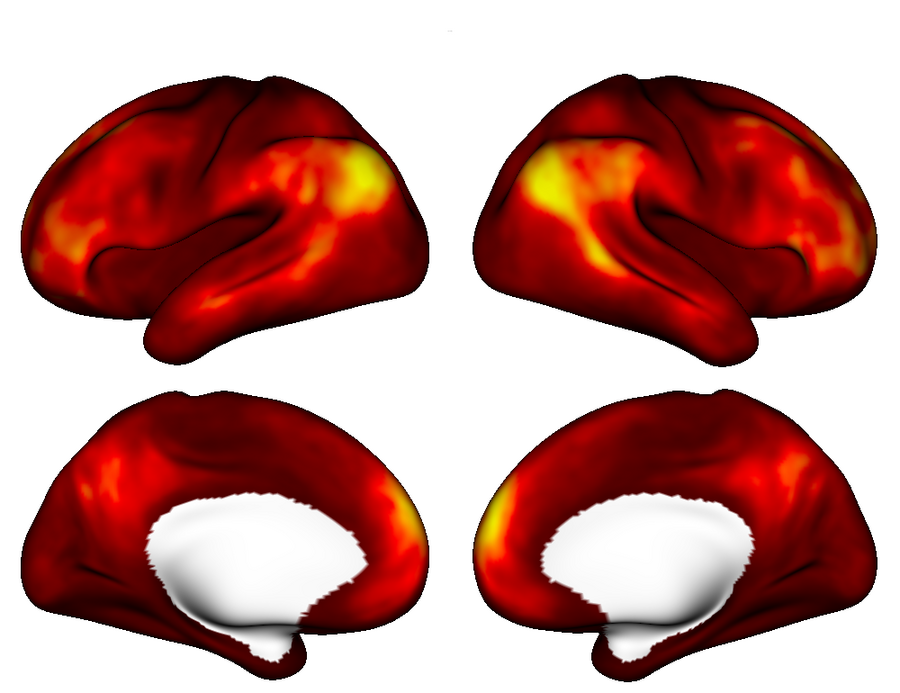} \\
\noalign{\vskip 2pt}\hline\noalign{\vskip 6pt}
\rotatebox{90}{\makecell[c]{\textbf{GICA15}\\($Q = 15$)}} &
\includegraphics[width=\linewidth]{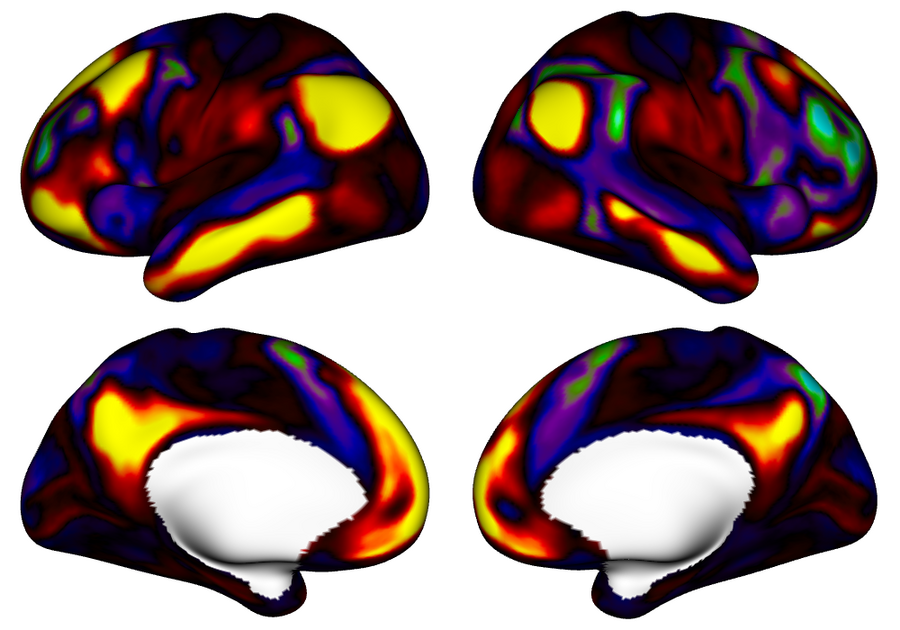} &
\includegraphics[width=\linewidth]{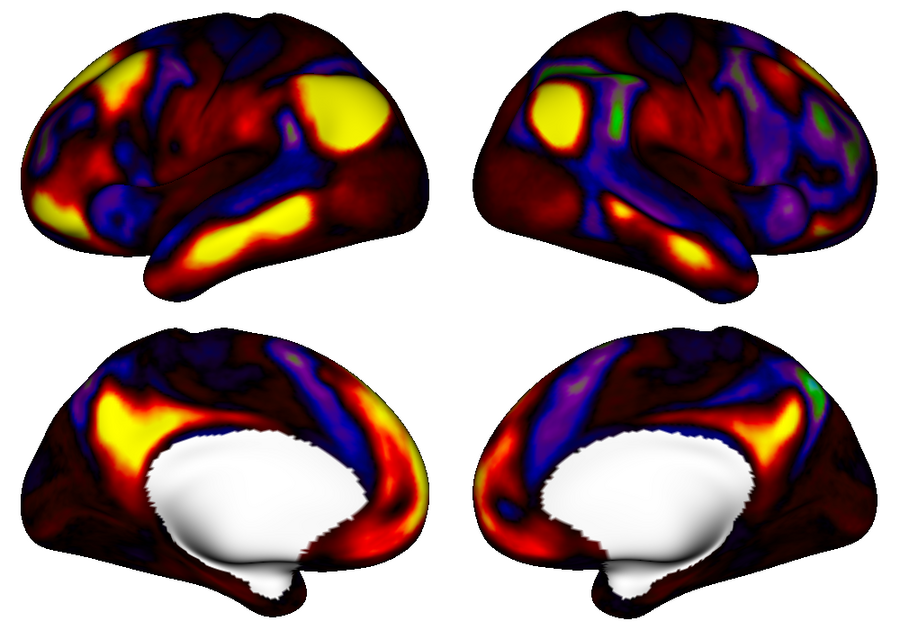} &
\includegraphics[width=\linewidth]{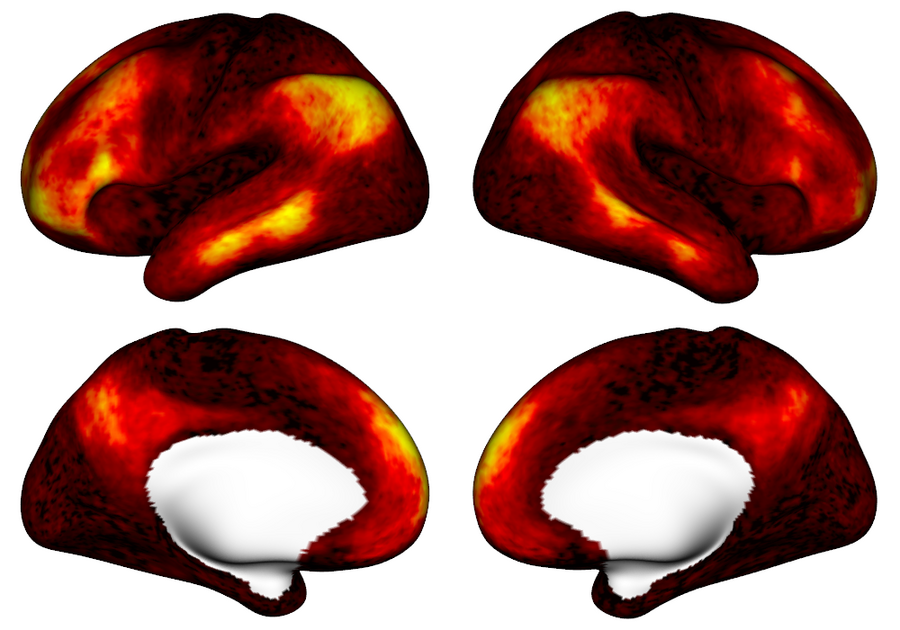} \\
\noalign{\vskip 2pt}\hline\noalign{\vskip 6pt}
\rotatebox{90}{\makecell[c]{\textbf{Yeo17}\\($Q = 17$)}} &
\includegraphics[width=\linewidth]{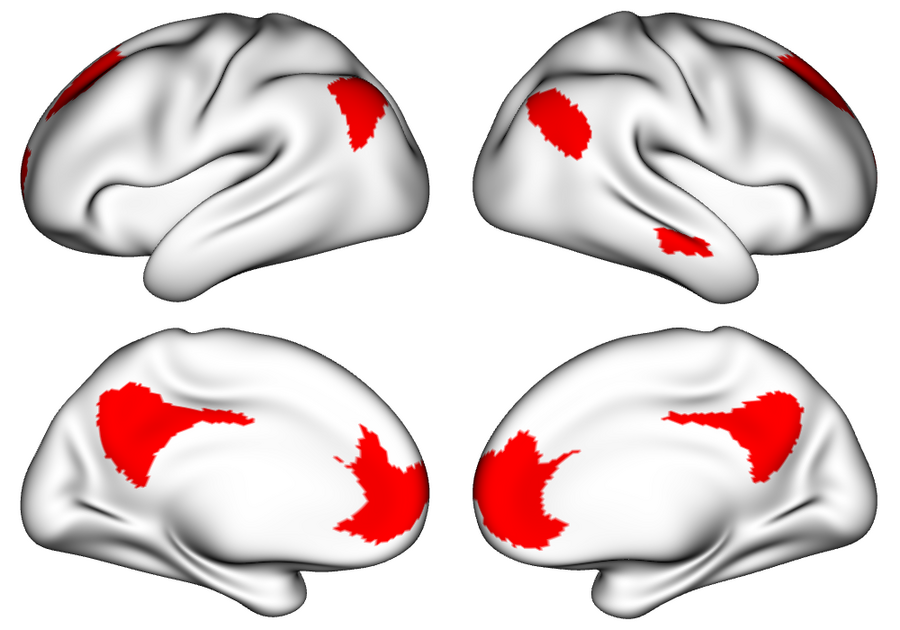} &
\includegraphics[width=\linewidth]{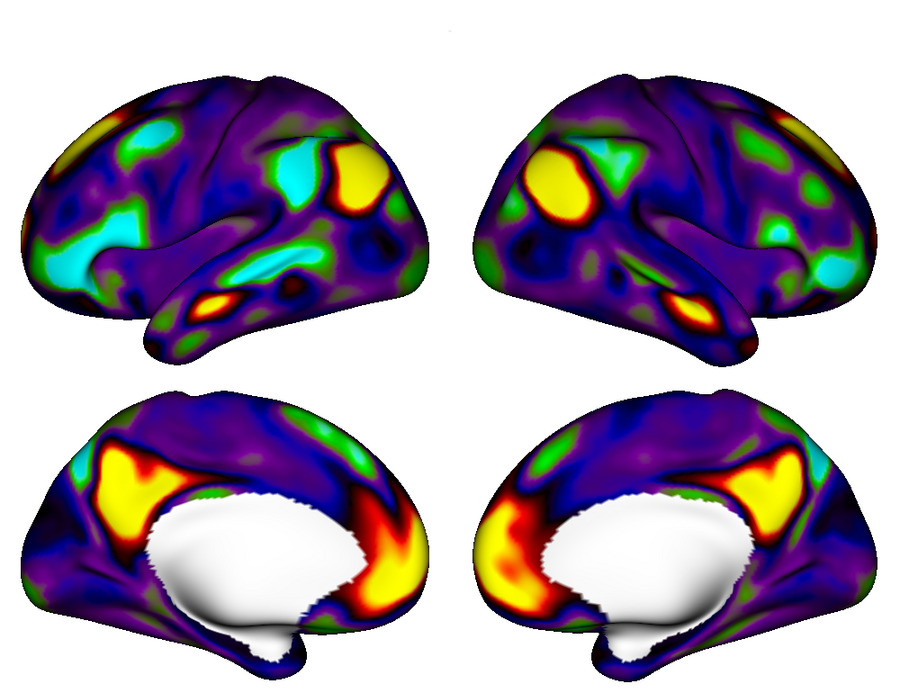} &
\includegraphics[width=\linewidth]{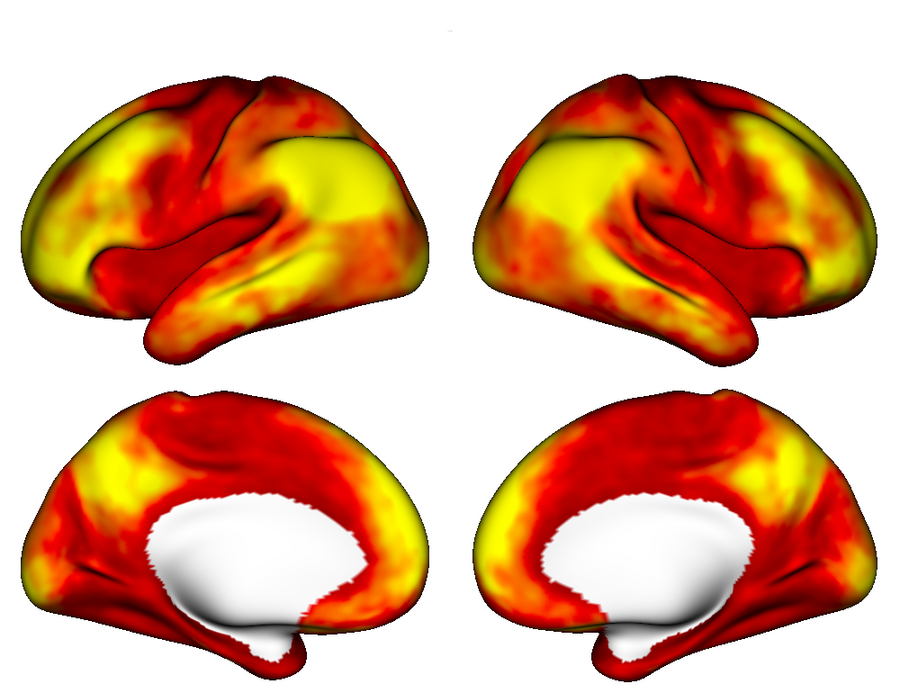} \\
\noalign{\vskip 2pt}\hline\noalign{\vskip 6pt}
\rotatebox{90}{\makecell[c]{\textbf{MSC}\\($Q = 17$)}} &
\includegraphics[width=\linewidth]{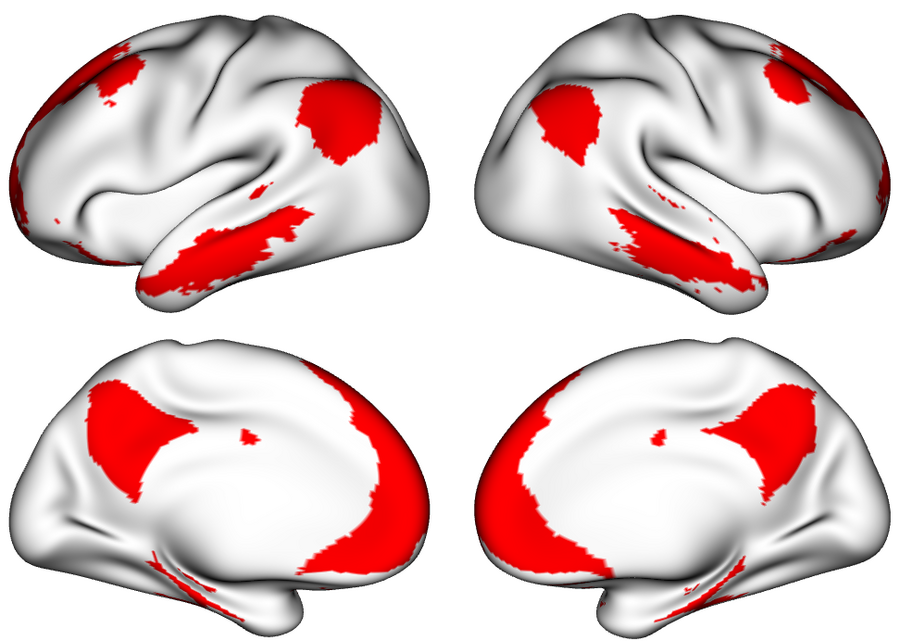} &
\includegraphics[width=\linewidth]{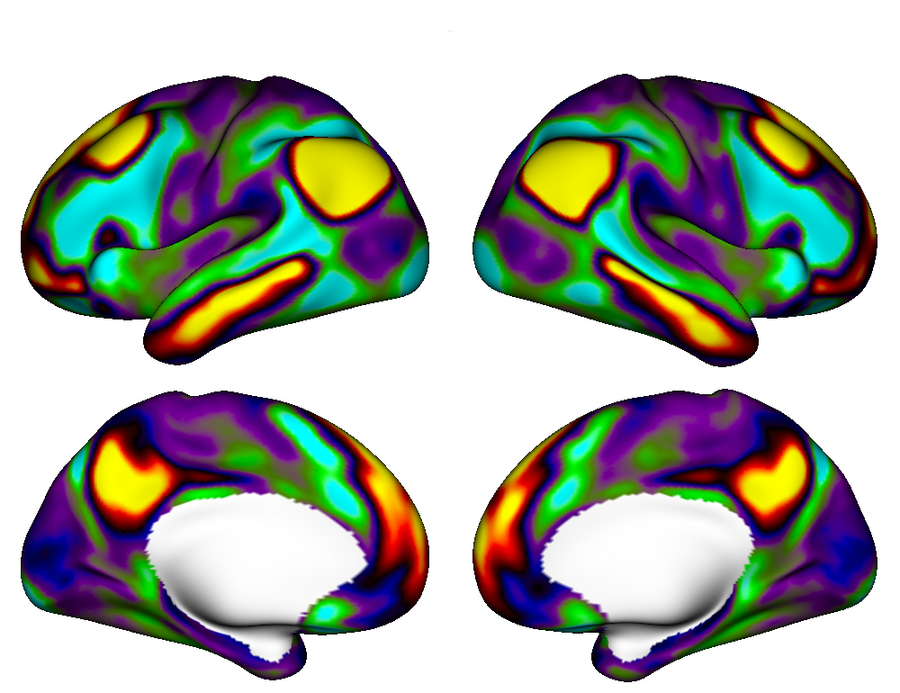} &
\includegraphics[width=\linewidth]{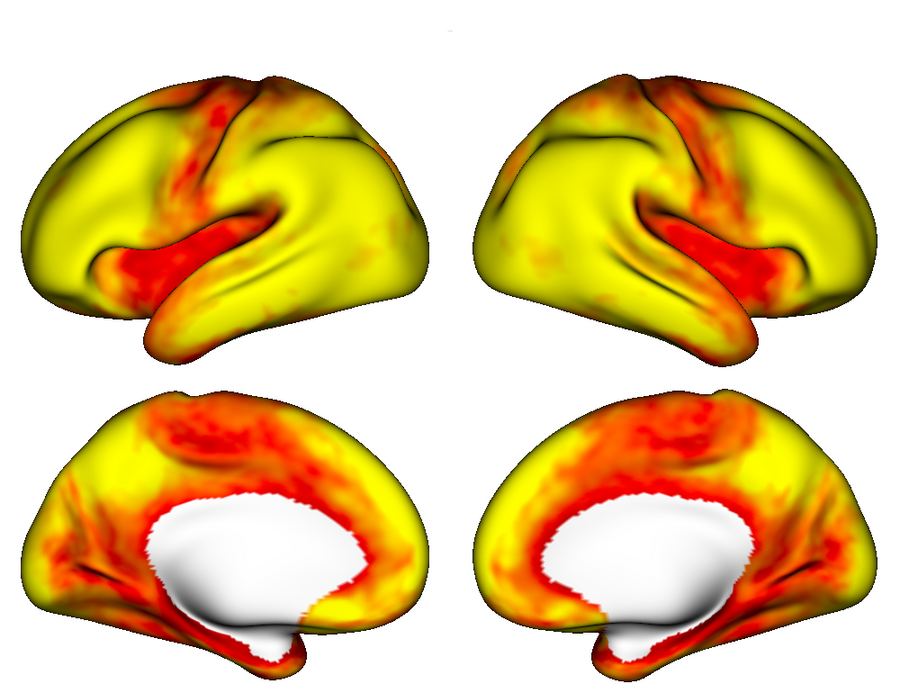} \\
\noalign{\vskip 2pt}\hline\noalign{\vskip 6pt}
\rotatebox{90}{\makecell[c]{\textbf{GICA25}\\($Q = 25$)}} &
\includegraphics[width=\linewidth]{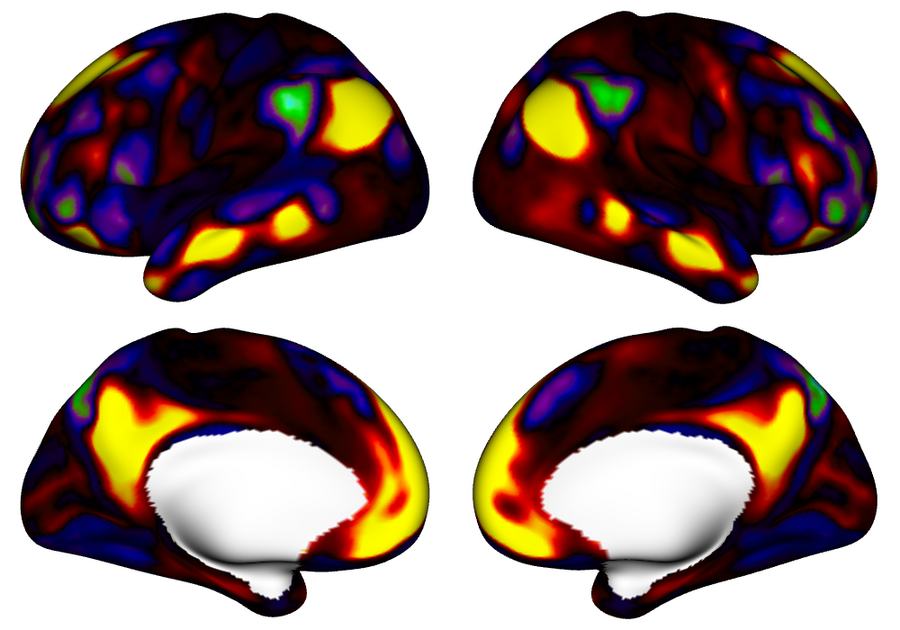} &
\includegraphics[width=\linewidth]{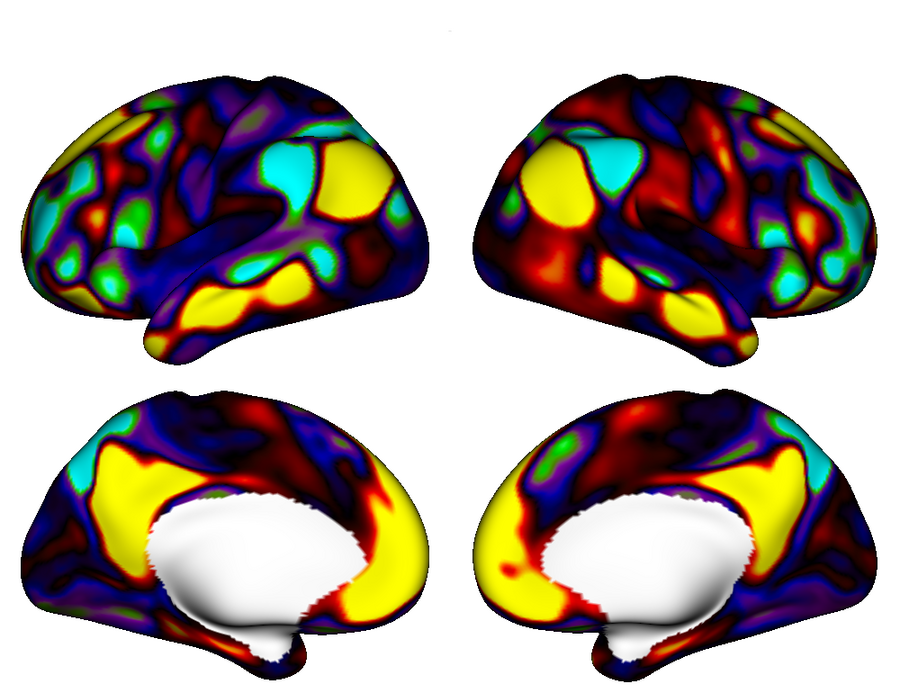} &
\includegraphics[width=\linewidth]{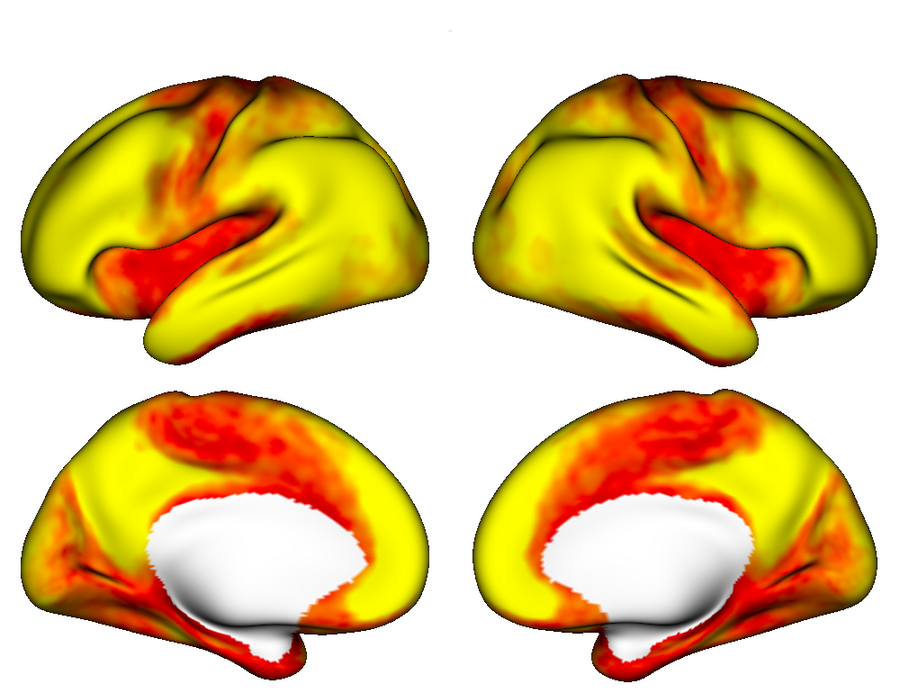} \\
\noalign{\vskip 2pt}\hline\noalign{\vskip 6pt} \\
&
% \rotatebox{90}{\makecell[c]{\textbf{GICA50}\\(IC 12)}} &
% \includegraphics[width=\linewidth]{Figure2/GICA50_IC12.png} &
% \includegraphics[width=\linewidth]{Figure2/GICA50_IC12_mean.png} &
% \includegraphics[width=\linewidth]{Figure2/GICA50_IC12_sd.png} \\
% \noalign{\vskip 2pt}\hline\noalign{\vskip 6pt}

\includegraphics[width=0.8\linewidth]{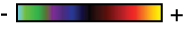} &
\includegraphics[width=0.8\linewidth]{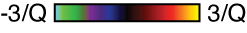} &
\includegraphics[width=0.8\linewidth]{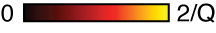} \\
\end{tabular}
\caption{\textbf{HCP-derived spatial topography priors for a default mode network component from various templates.} Templates include two parcellations (Yeo17 and MSC) and two types of network maps (ICA and PROFUMO) and are ordered from lowest to highest resolution (number of networks). The specific networks displayed are PROFUMO mode 8, GICA15 component 2, Yeo17 Default A, MSC Default, and GICA25 IC 2. Template maps are displayed on an arbitrary scale, while the scale of the priors is inversely proportional to the template resolution $Q$, given the additive nature of the BBM decomposition.}
\label{fig:HCP_priors}
\end{figure}

\subsection{Different templates produce similar network topography priors}

Spatial topography priors for a default mode network component from various templates are shown in Figure \ref{fig:HCP_priors}. Similar features are seen across the different priors, even though the templates are based on distinct model assumptions. The prior variance maps show that individual variation in engagement is greatest in areas of high average engagement, reflecting differences in the configuration, size, and distribution of networks across individuals. \review{Note that the prior variance reflects differences in engagement intensity, not simply network presence.} The low prior variance in background areas, where true population variability is low, will result in a high degree of regularization in these areas for the corresponding networks. Higher prior variance in areas of engagement, by contrast, will allow for individual features to be expressed. Furthermore, due to the additive nature of the decomposition, noise reduction in background areas of certain networks will indirectly improve estimation of engagement in other networks corresponding to the same physical location.

% Figure 4 shows the impact of sample size

%macros for template and prior mean maps for Fig 2. Change origin directory for Figure2_Yeo17 for different colors for each network.
\renewcommand{\PictureA}[1]{
    \includegraphics[width=6cm, trim = 0 14.5cm 0 3cm, clip]{Figure_samplesize/#1}
}
\renewcommand{\PictureB}[1]{
    \includegraphics[width=6cm, trim = 0 14.5cm 0 3cm, clip]{Figure_samplesize/#1} 
}

% I do not understand why this is needed because the arrow pictures are the exact same as Picture B
\renewcommand{\PictureC}[1]{
    \includegraphics[width=6cm, trim = 0 11.5cm 0 1.5cm, clip]{Figure_samplesize/#1} 
}
%macro for row labels 
\renewcommand{\RotLabel}[1]{
  \begin{picture}(0,60)\put(0,30){\rotatebox[origin=c]{90}{#1}}\end{picture}
}

\begin{figure}
\centering
\small
\begin{tabular}{ccc}
 & \textbf{Prior Mean} & \textbf{Prior SD} \\[10pt]
 \hline\\[-6pt]
{\RotLabel{$n=25$}} &
{\PictureA{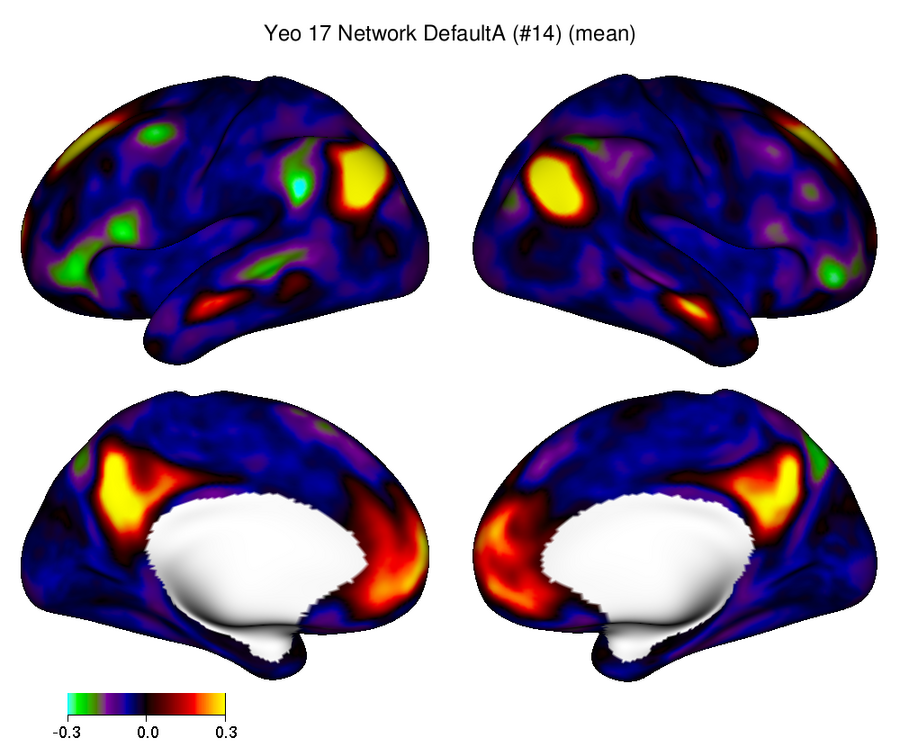}} &
\PictureB{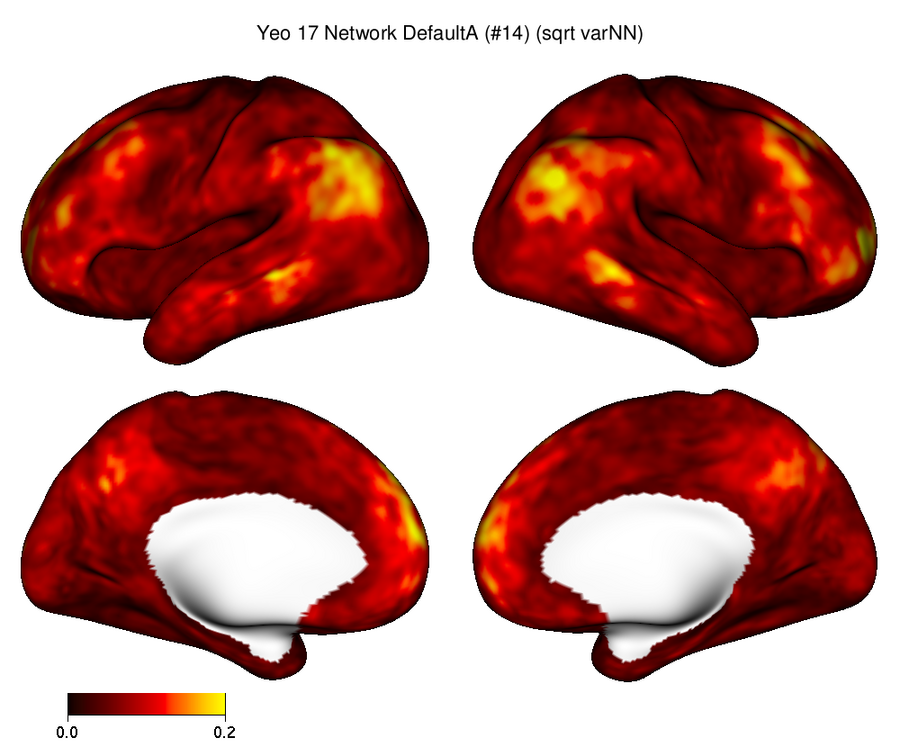} \\
\hline\\[-6pt]
{\RotLabel{$n=50$}} &
{\PictureA{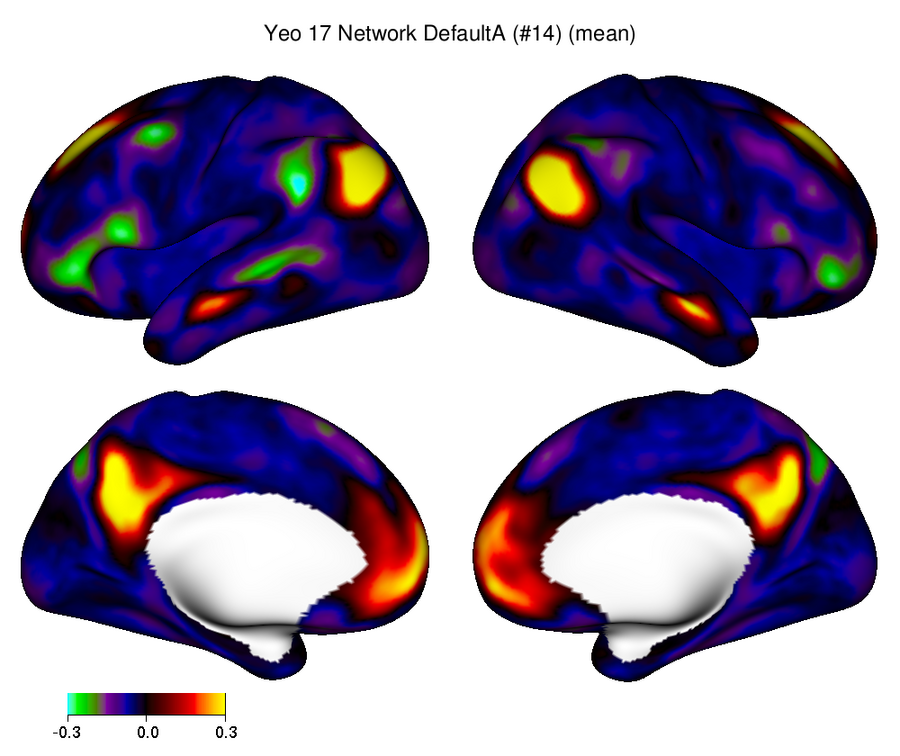}} &
\PictureB{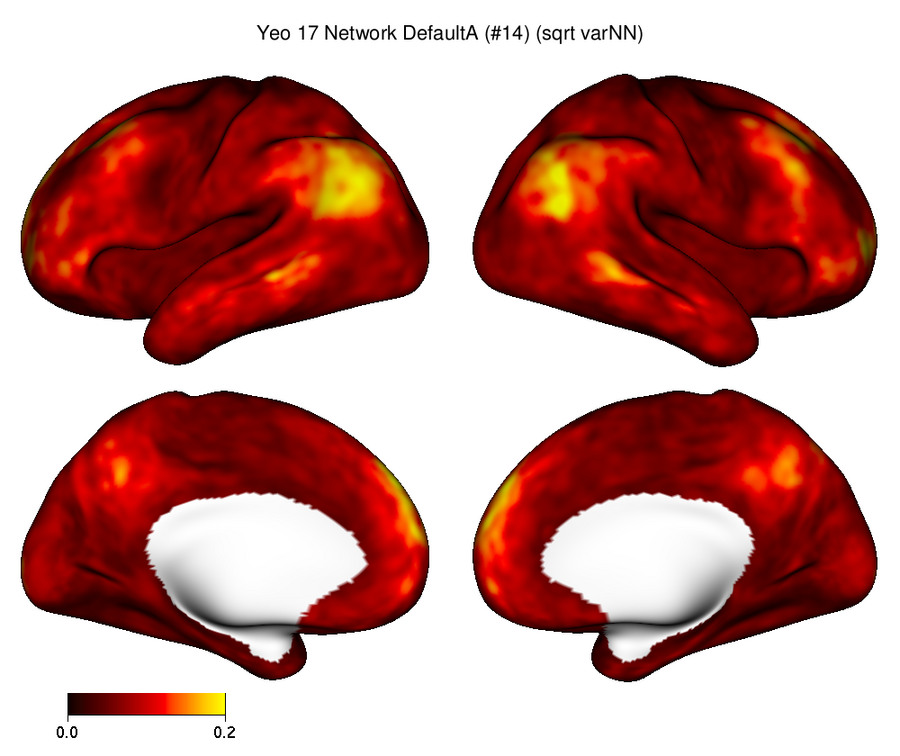} \\
 \hline\\[-6pt]
{\RotLabel{$n=100$}} &
{\PictureA{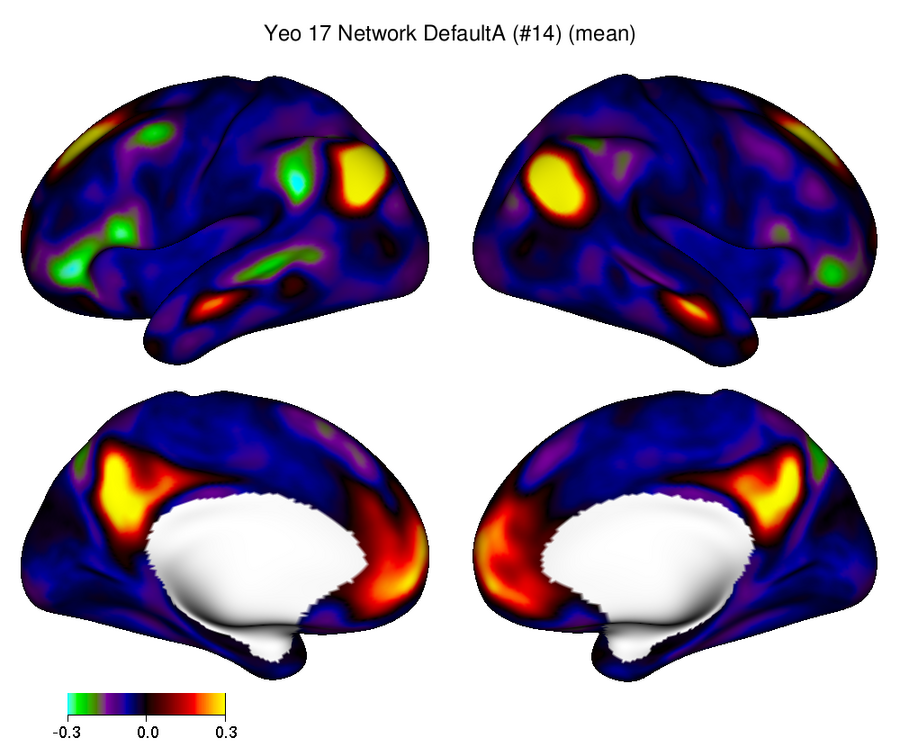}} &
\PictureB{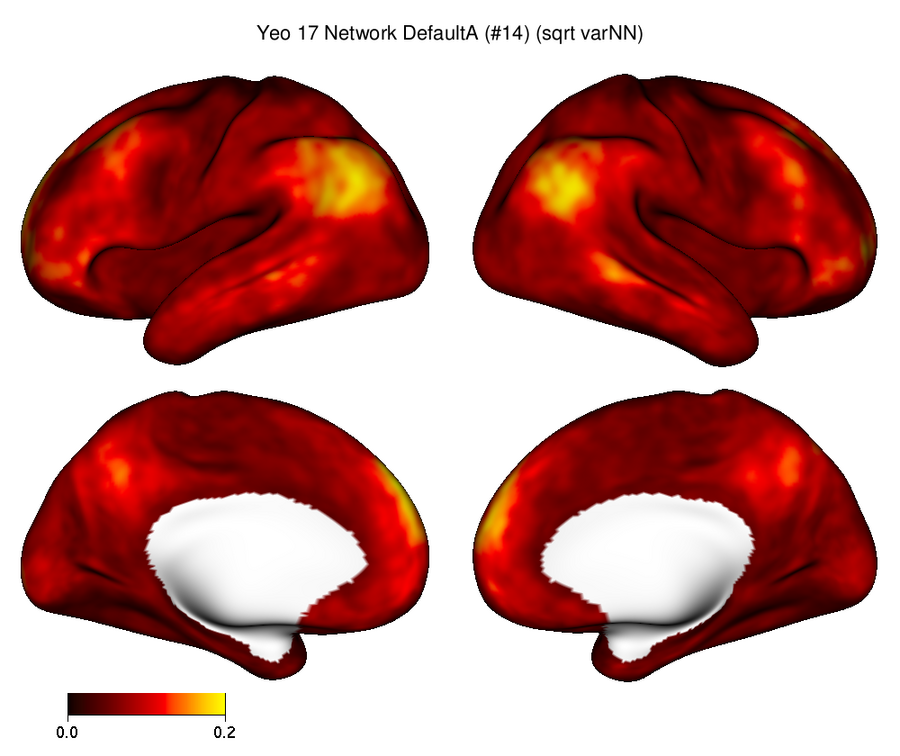} \\
 \hline\\[-6pt]
{\RotLabel{$n=200$}} &
{\PictureA{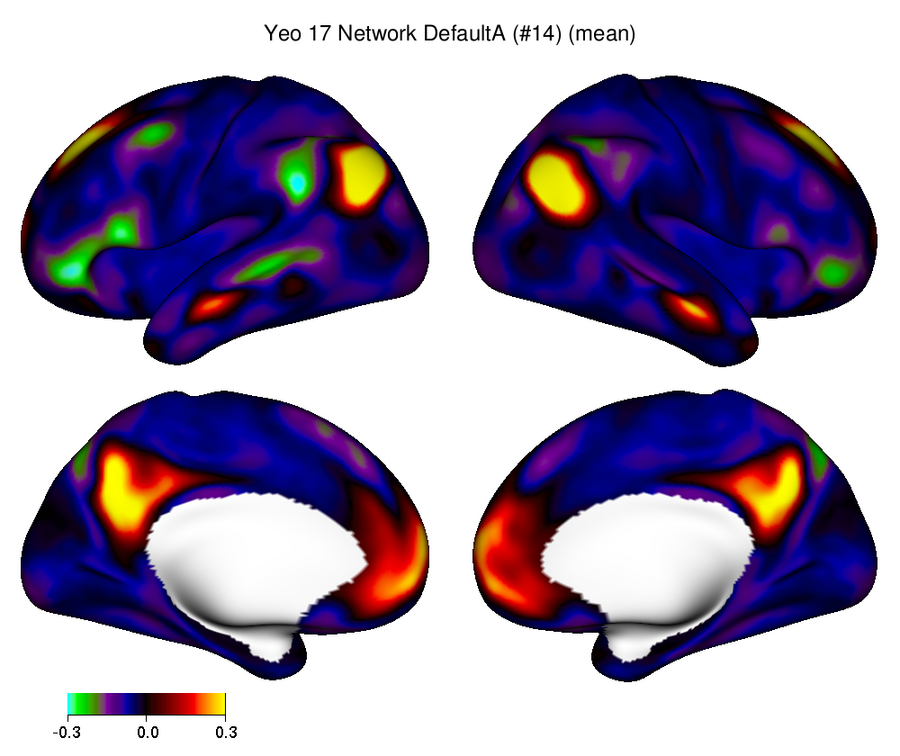}} &
\PictureB{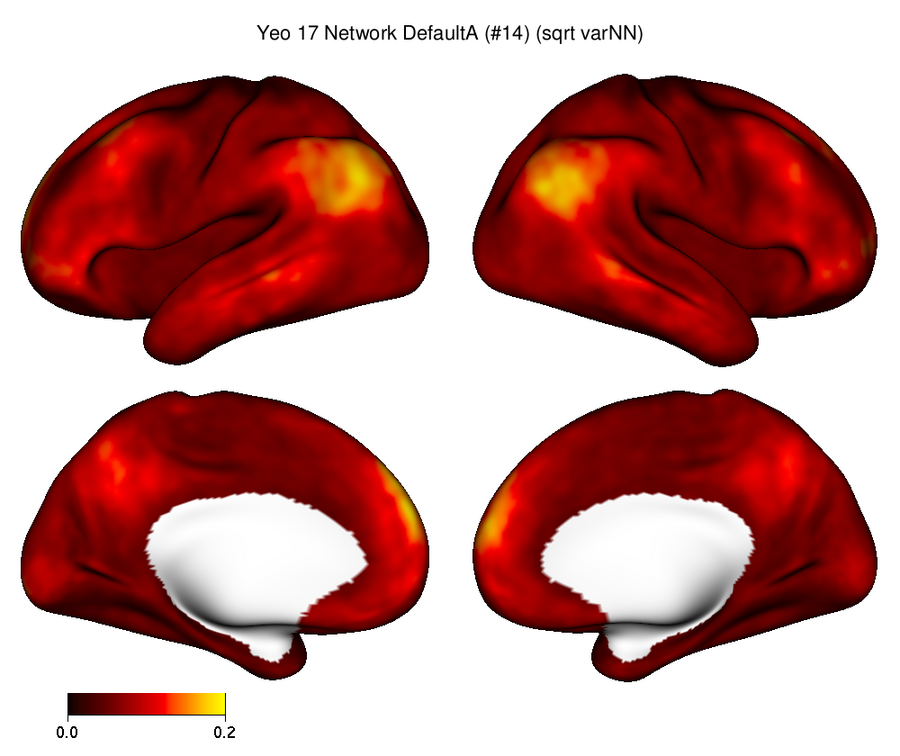} \\
\hline\\[-6pt]
{\RotLabel{$n=348$}} &
{\PictureA{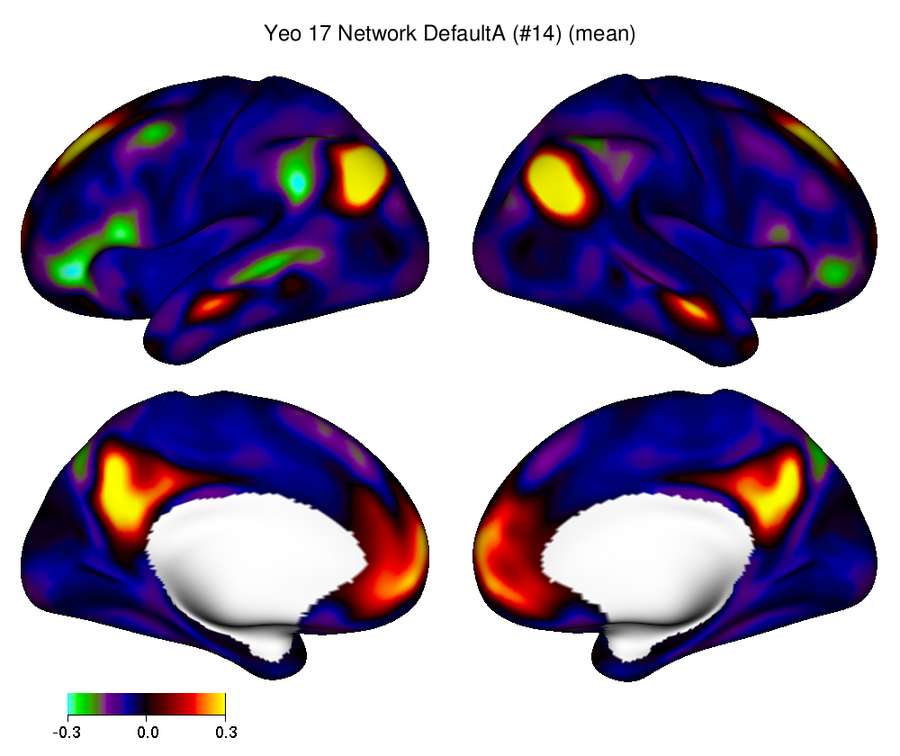}} &
\PictureB{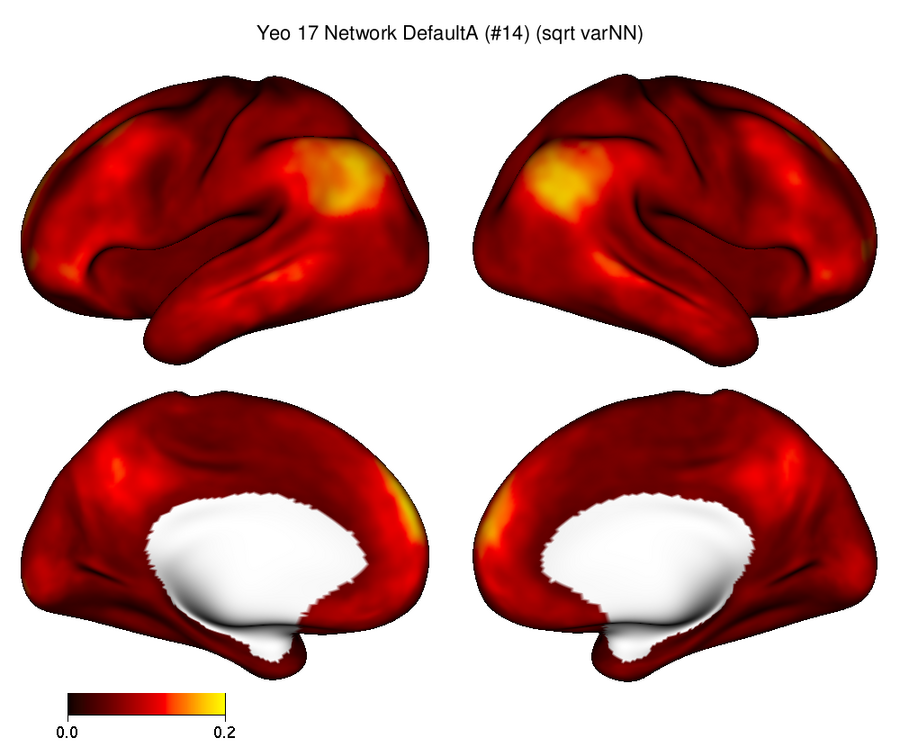} \\
\hline\\[-6pt]
\end{tabular}
\caption{\review{\textbf{Effect of sample size on spatial priors.} Default mode network A from the Yeo17 template. The prior mean is highly stable even with $n=25$ training participants, while the prior standard deviation is more sensitive to sample size. }}
\label{fig:samplesize_yeo17}
\end{figure}

% Figure 4 ssecond version includes smoothing 

%macros for template and prior mean maps for Fig 2. Change origin directory for Figure2_Yeo17 for different colors for each network.
\renewcommand{\PictureA}[1]{
    \includegraphics[width=3cm, trim = 17cm 14.5cm 0 3cm, clip]{Figure_samplesize/#1}
}
\renewcommand{\PictureB}[1]{
    \includegraphics[width=3cm, trim = 17cm 14.5cm 0 3cm, clip]{Figure_samplesize/#1} 
}

% I do not understand why this is needed because the arrow pictures are the exact same as Picture B
\renewcommand{\PictureC}[1]{
    \includegraphics[width=6cm, trim = 0 11.5cm 0 1.5cm, clip]{Figure_samplesize/#1} 
}
%macro for row labels 
\renewcommand{\RotLabel}[1]{
  \begin{picture}(0,60)\put(0,30){\rotatebox[origin=c]{90}{#1}}\end{picture}
}

\newcommand{\PictureAo}[1]{
    \includegraphics[width=4.75cm, trim = 0cm 28mm 2cm 2cm, clip]{Figure_FC/#1}
}

\renewcommand{\PictureA}[1]{
    \includegraphics[width=4cm, trim = 3cm 28mm 2cm 2cm, clip]{Figure_FC/#1}
}

\newcommand{\PictureBo}[1]{
    \includegraphics[width=4.75cm, trim = 0cm 0cm 2cm 2cm, clip]{Figure_FC/#1}
}

\renewcommand{\PictureB}[1]{
    \includegraphics[width=4cm, trim = 3cm 0cm 2cm 2cm, clip]{Figure_FC/#1}
}

\begin{figure}
\centering
\begin{tabular}{C{0.03\textwidth}C{0.3\textwidth}C{0.26\textwidth}C{0.26\textwidth}}
 & {\hspace{0.75cm}\textbf{Population}} & {\textbf{Cholesky}} & {\textbf{Inverse-Wishart}} \\[10pt]
\rotatebox{90}{{Mean}} &
\PictureAo{prior_combined_Yeo17_noGSR_FC_Empirical_mean.png} &
\PictureA{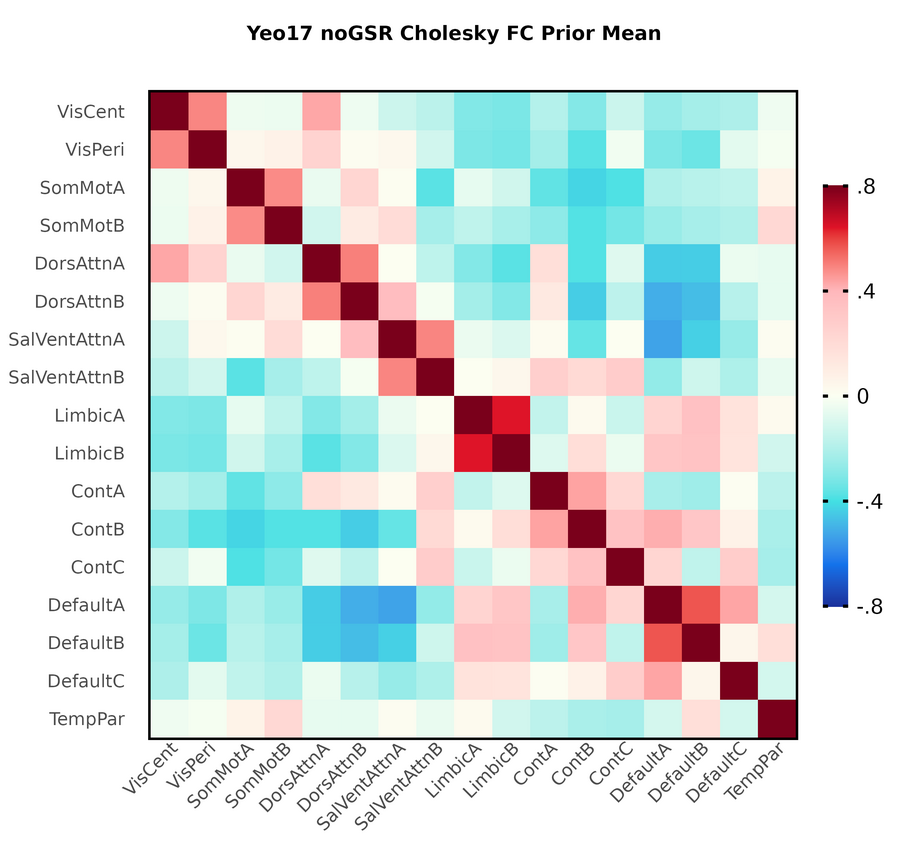} &
\PictureA{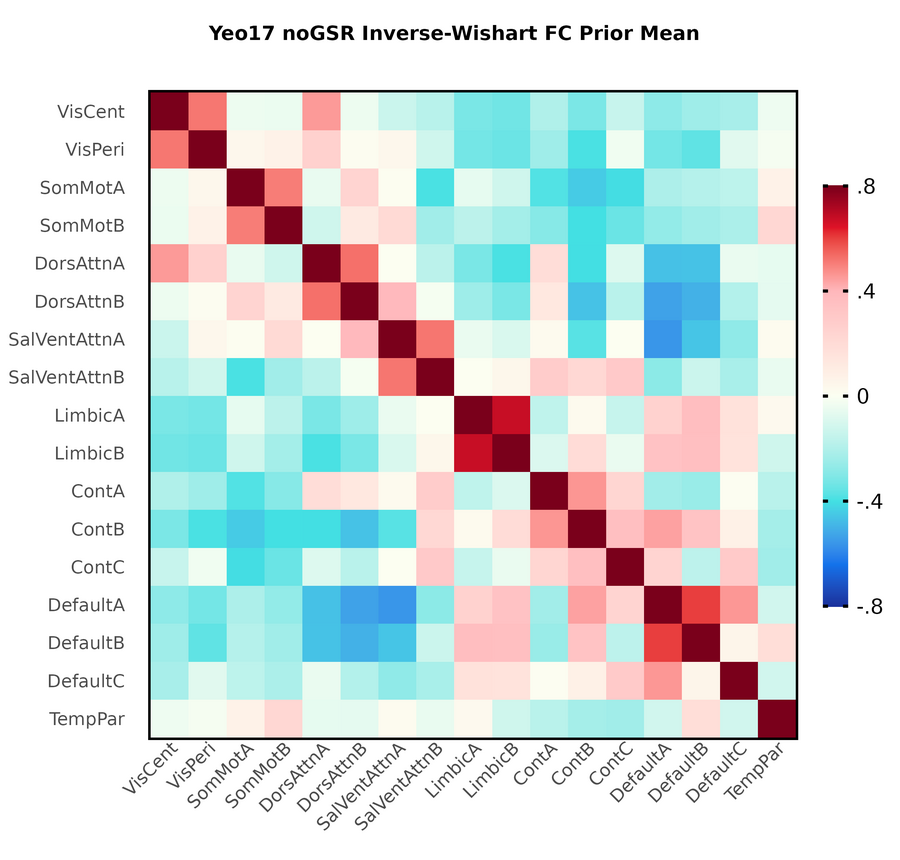} \\
\\[-4pt]
\rotatebox{90}{{SD}} &
\PictureBo{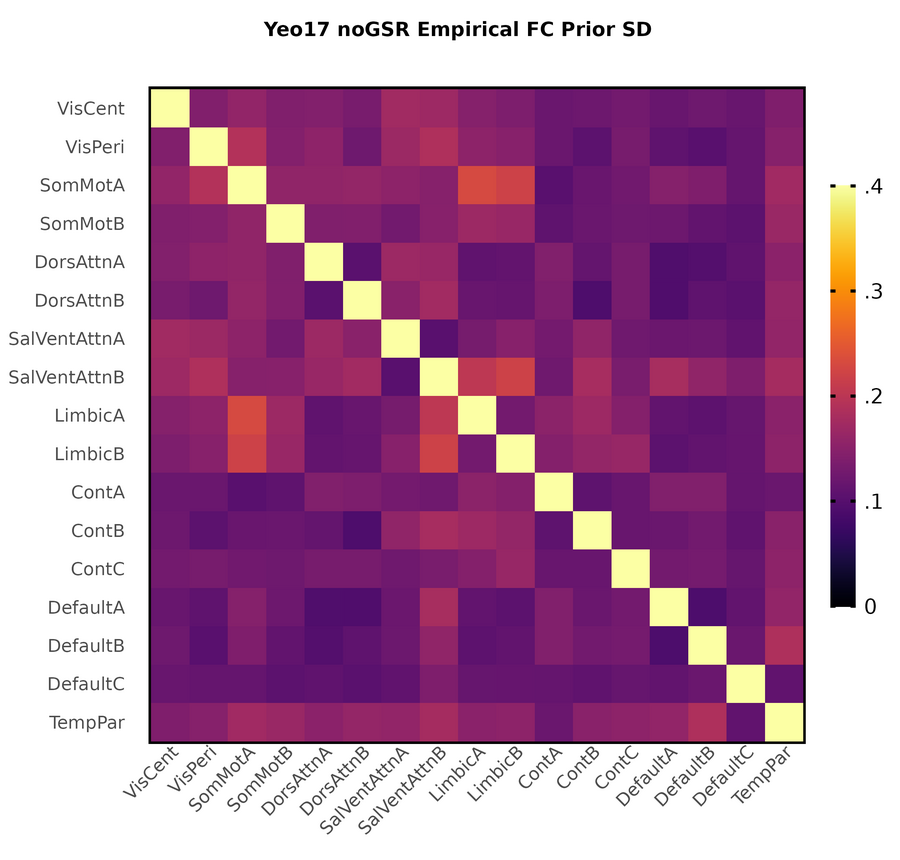} &
\PictureB{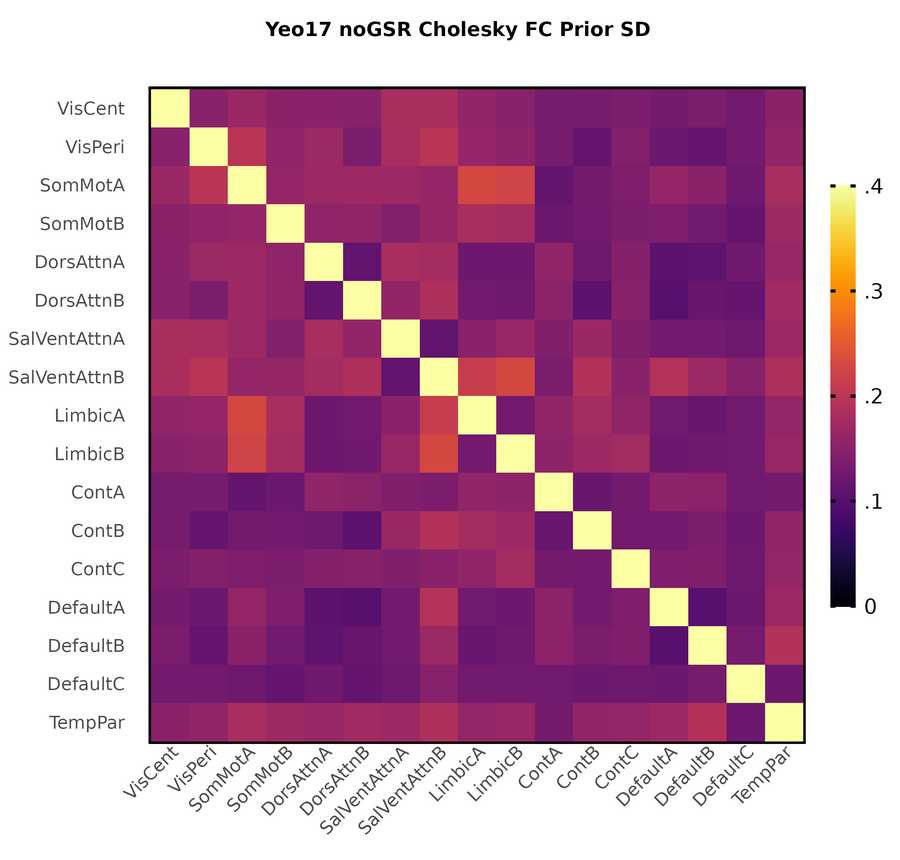} &
\PictureB{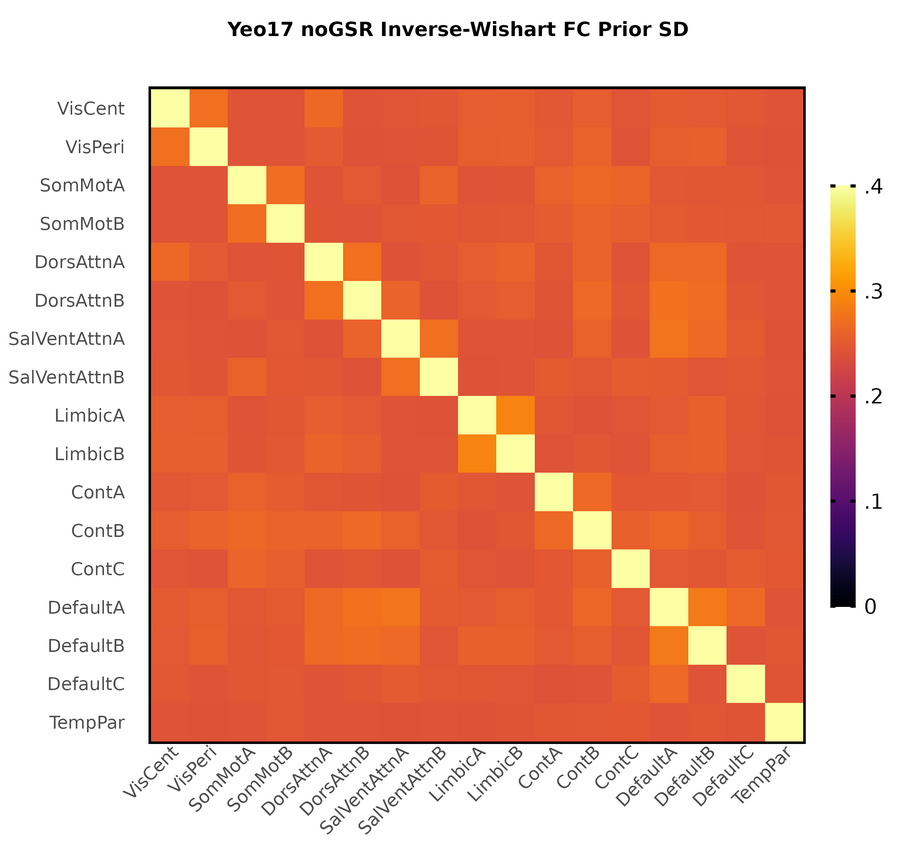} \\
\end{tabular}
\caption{\textbf{Example of the BBM functional connectivity (FC) prior.} Results shown here correspond to the Yeo17 template; results for other templates are displayed in Appendix Figure \ref{fig:FC_empirical}. The first column shows the element-wise empirical mean and standard deviation (SD) within the training set, which is used to establish the priors. The second and third columns show the element-wise mean and SD based on the Cholesky and the inverse-Wishart priors. The population mean is captured by both priors. However, the inverse-Wishart prior fails to capture the element-wise population variance patterns, given that distribution's limited flexibility. The Cholesky prior, which we developed for BBM, accurately captures the element-wise population variance patterns.}
\label{fig:FC_Yeo17}
\end{figure}

\subsubsection{BBM FC prior accurately captures population variance patterns}

The FC priors in BBM are illustrated in Figure \ref{fig:FC_Yeo17} for the Yeo17 template and in Appendix Figure \ref{fig:FC_empirical} for other templates.  The element-wise empirical mean and standard deviation within the training set, representing population tendencies and variability, are shown in the first column. The second two columns show the corresponding values based on the two choices of FC prior in BBM: the permuted Cholesky prior described in \cite{mejia2025leveraging} and the conjugate inverse-Wishart (IW), which is also used in PROFUMO \citep{harrison2020modelling}. While both priors capture the population mean, only the Cholesky prior accurately captures the population variance. This is because the IW distribution for covariance matrices includes a single scalar parameter that moderates the variance across the matrix, so that the element-wise variance is monotonically related to the element-wise mean. In BBM, this parameter is chosen to minimize the difference between the population and prior element-wise variance, but with a constraint that the prior variance should not be less than the population variance anywhere. This avoids an overly informative prior, which would result in too much shrinkage towards the group and loss of relevant individual information. Due to this limitation of the inverse-Wishart distribution, the IW FC prior tends to have much higher variance for many FC edges compared with the population, resulting in unnecessarily weak shrinkage. The Cholesky prior, by contrast, is more informative while not being overly informative. This theoretically maximizes accuracy of FC estimates, and has indeed been shown to produce more reliable and individualized estimates of between-network FC \citep{mejia2025leveraging}. Improved estimation of the temporal component of the BBM model may also have downstream benefits for estimation of spatial topography, since the temporal and spatial components are estimated iteratively.

% ----------------------------------------------------
% Figure 7 -- Example HCP subject
% ----------------------------------------------------

\begin{figure}
\centering
\small
% -------- Top row (cc only) --------
\begin{tabular}{cc}
{Prior Mean} & {Prior SD}\\[10pt]
\includegraphics[width=4cm]{Figure2/Yeo17_IC14_mean.png} &
\includegraphics[width=4cm]{Figure2/Yeo17_IC14_sd.png} \\
\includegraphics[width=3cm]{Figure2/legend_mean.png} &
\includegraphics[width=3cm]{Figure2/legend_sd.png} \\[-2pt]
\end{tabular} 

\vspace{12pt}

% -------- Bottom block (original layout: cccl) --------
\begin{tabular}{cccl}
{Posterior Mean} & {Posterior SD} & {Engagements} \\[10pt]
\includegraphics[width=4cm]{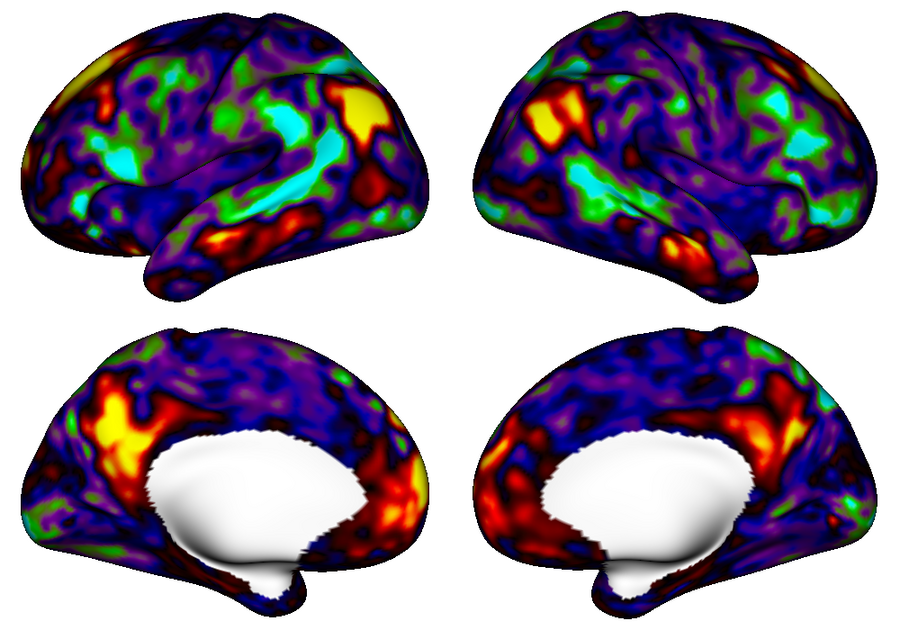} &
\includegraphics[width=4cm]{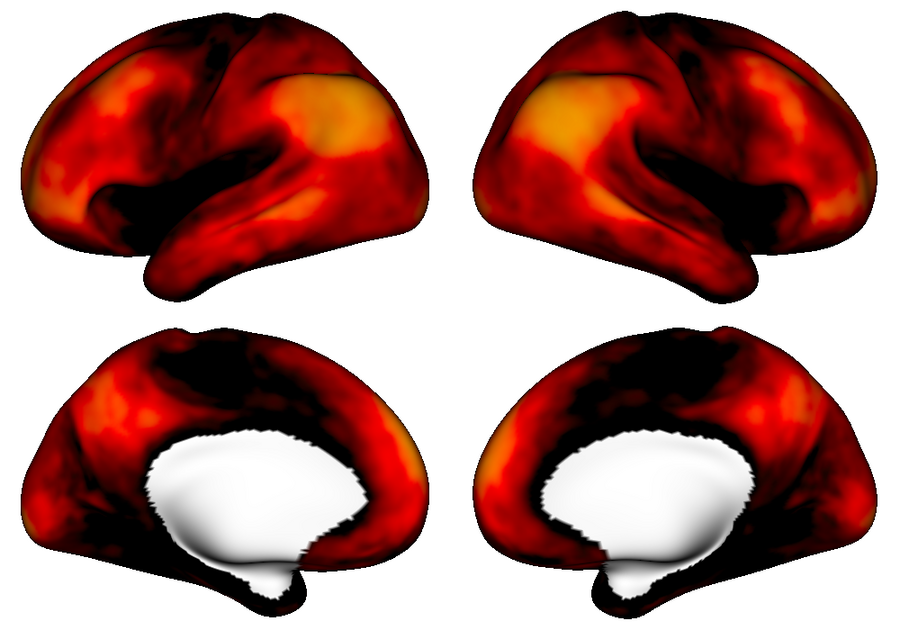} &
\includegraphics[width=4cm]{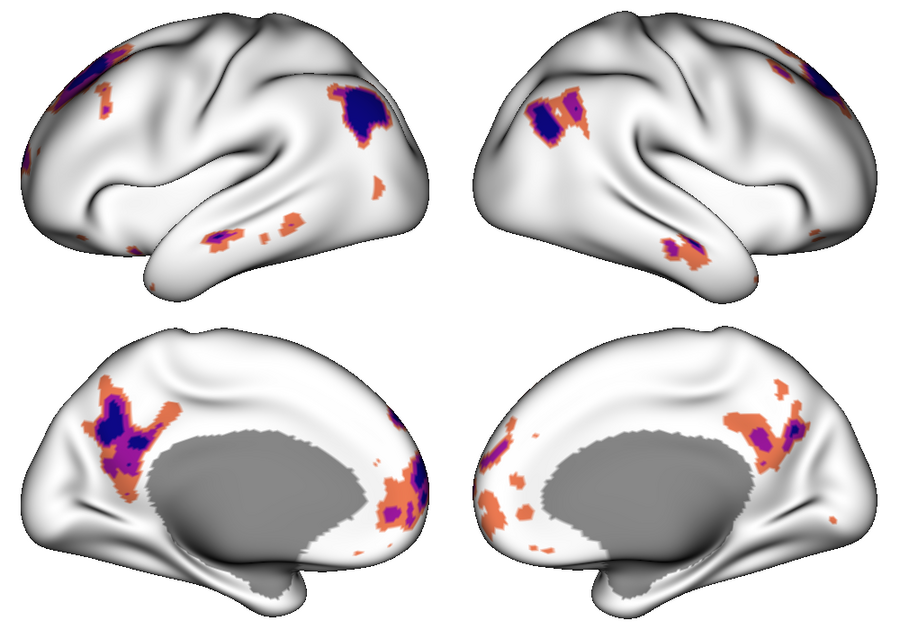} &
\includegraphics[width=25mm, trim = 5mm 0 0 18mm, clip]{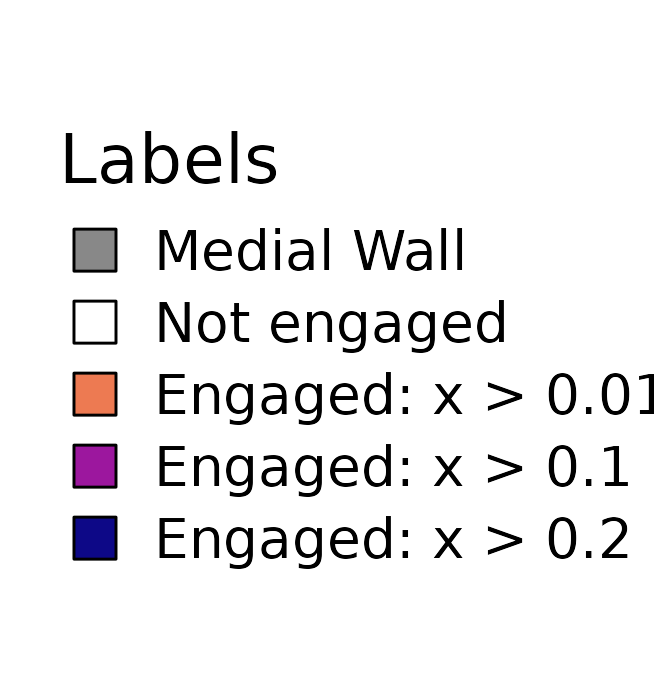} \\
\includegraphics[width=3cm]{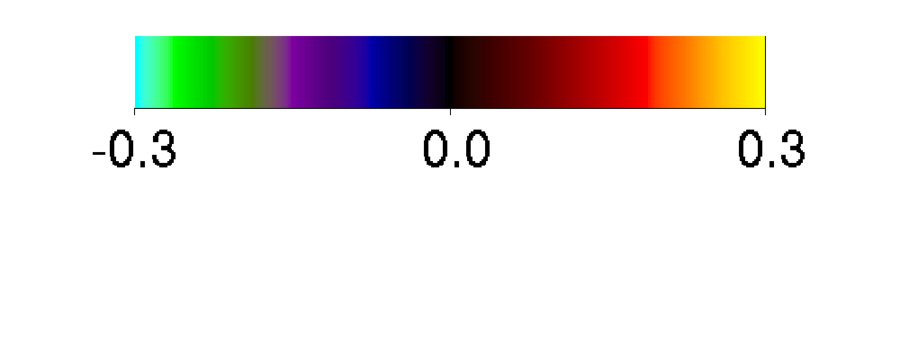} &
\includegraphics[width=3cm]{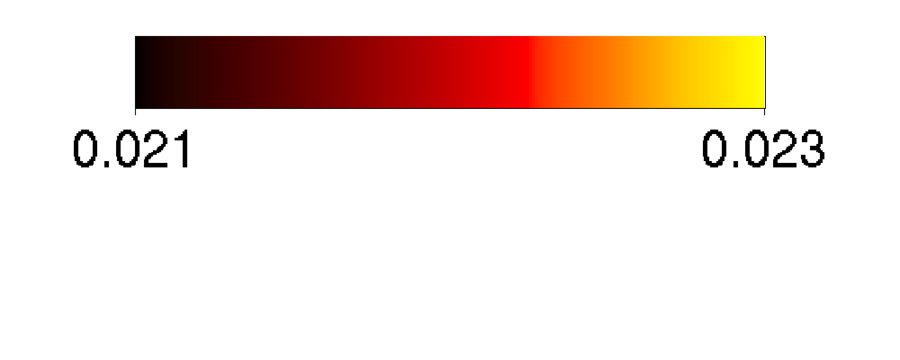} \\[-2pt]
\end{tabular}
\caption{\textbf{Example individual BBM spatial engagement maps.} A single HCP subject was analyzed using BBM with population-derived priors based on the Yeo17 template. The BBM posterior mean and standard deviation for a network corresponding to the Yeo17 DefaultA parcel are shown, along with areas of statistically significant engagement. Significance is based on the Bayesian equivalent of a hypothesis test with $\alpha = 0.01$ and Bonferroni correction within each network. Engagements are defined in percentage signal change with respect to 1\% change in the temporal fluctuation of the signal over its mean value, shown at 0.01\%, 0.1\%, and 0.2\% thresholds. This is analogous to the common practice of thresholding group ICA maps at a certain number of standard deviations from the mean to isolate the main areas of engagement.}
\label{fig:HCP_subject_Yeo17}
\end{figure}

\subsection{Illustration of BBM for personalized network topography and inference}

Figure \ref{fig:HCP_subject_Yeo17} \review{shows an example of} personalized functional network topography maps produced by applying BBM to analyze data from one HCP subject, using the population-derived priors based on the Yeo17 template \review{(top row). The individual-level maps (bottom row) represent the posterior distribution for $s_{qv}$, the subject-specific network engagement in the BBM model in equation (\ref{eqn:BBM_model}).}  A single network is shown corresponding to the Default A parcel. The posterior mean provides an estimate of engagement in the network at every vertex, while the posterior standard deviation represents uncertainty around that estimate. Binary maps of statistically significant engagement are produced based on the posterior mean and standard deviation, using the Bayesian equivalent of a hypothesis test. A range of ``null hypotheses'' based on different minimum effect sizes of interest are tested against, resulting in nested sets of significantly engaged vertices. The ability to specify a minimum effect size gives the researcher flexibility to focus on locations that exhibit intense network engagement, or to consider all vertices that exhibit non-zero engagement. The researcher can also consider a range of minimum effect sizes, as we have done here, to distinguish between levels of engagement.

\review{
To quantify the effect of the BBM model on the posterior, we estimated the degree of shrinkage at every location within each network, and examined the level of shrinkage as a function of network engagement level. Specifically, we defined shrinkage for network $q$ at location $v$ as the distance from the maximum likelihood estimate (MLE) to the posterior mean, divided by the distance from the MLE to the prior mean:}

\review{\begin{equation}
\lambda_q(v) = \frac{|\hat{s}^{MLE}_q(v)-\hat{s}^{BBM}_q(v)|}{|\hat{s}^{MLE}_q(v) - s_q^0(v)|},
\label{eq:lambda}    
\end{equation}}

\review{where $\hat{s}^{BBM}_q(v)$ is the BBM posterior mean produced, $\hat{s}^{MLE}_q(v)$ is the MLE (produced by regressing the network time series against the BOLD data), and $s_q^0(v)$ is the prior mean.}
\review{$\lambda_q(v)$ was calculated for 10 randomly selected HCP participants, using two runs per subject. The degree of shrinkage was then plotted as a function of posterior mean value for each network (Figure \ref{fig:app:shrinkage}). This reveals that shrinkage is high in background regions with posterior mean close to zero (40-50\% on average), while in areas of moderate to high engagement shrinkage is low (approximately 10\% on average).} 

\subsection{Computational demands of BBM}
Prior estimation based on the templates listed in Table \ref{tab:template-summary} was completed for the HCP population of 348 participants in an average of 155 minutes per template, with peak memory usage of 126 GB. All calculations were performed using 48 concurrent threads on an Intel Xeon W 2.7 GHz processor running macOS Tahoe 26.2. The number of participants in the sample, the number of volumes in each scan, and the number of vertices will directly affect the computational demands of \texttt{estimate\_prior()}. It is worth noting that this calculation needs to be performed only once for each training dataset. For representative populations, priors can ideally be shared through a repository (OSF in the present study), thereby eliminating the need for repeated prior estimations and allowing users to proceed directly to the \texttt{fit\_BBM()} function, which fits the model to individual data for the focal subject(s) of the study.
\review{Fitting BBM with the IW FC priors to a single HCP session required 7 minutes on a single thread and used 4.3 GB of memory. %Using the Cholesky FC priors demanded 23 hours and 11 GB of memory, illustrating the substantially higher computational burden posed by the Cholesky FC method. 
This process can be parallelized across participants within a study, thus reducing the time required to analyze an entire study cohort.}

% TO DO: Report computational demands 

% \subsection{Computational demands of BBM}

% Report computation times for prior estimation.

% Report computation times for model fitting on HCP and MSC (per session).

% Report the computational resources used (processors, memory)

\section{Discussion}

% Restate the results
In this paper we have described and illustrated Bayesian brain mapping (BBM), a personalized functional brain network mapping technique that leverages population information through population-derived priors for highly reliable network topography and between-network FC estimation.  Those priors use established network maps or parcellations as templates, producing individual-specific analogues of known networks. The priors relax constraints of the template, allowing for overlap between networks and differential engagement within networks. The powerful noise reduction properties of the priors facilitate reliable estimation of individualized functional topography with moderate scan duration, mitigating the need for extensive scanning of individuals \citep{mejia2020template, derman2026thesis}. We also provide resources to facilitate its implementation, including a demo illustrating the \texttt{BayesBrainMap} R package functions for prior estimation, model fitting, and inference, as well as HCP-derived priors based on a range of templates.

% Importance of results
 \review{These resources and the generalizable BBM framework are aimed at lowering} the barrier to personalized functional brain network mapping by reducing required scan duration through noise-mitigating informative priors, while avoiding heavy computational demands associated with multi-subject hierarchical models. Furthermore, once the priors have been established, model estimation is fast and parallelizable. For experiments involving healthy adults, the HCP-derived priors we provide here can be applied directly, saving time and simplifying the analysis pipeline. For studies of other populations, our pipeline can be adapted to produce priors using other publicly available resources or data from the focal study. %By standardizing the FC priors and cortical surface space, this approach can also improve comparability across subjects and across studies.

 % Limitations of measures
Several limitations of the current BBM framework should be noted. First, priors should only be applied to study individuals coming from the same or similar populations. As such, the HCP-derived priors we share should not be utilized for the study of populations other than healthy young adults, such as children or the elderly. Other large publicly available datasets facilitate the use of BBM for certain populations including Alzheimer's disease \citep{ADNI}, adolescents \citep{casey2018adolescent}, infants \citep{edwards2022dhcp}, and many others. However, for studies of rarer populations it may be difficult to achieve samples large enough to construct accurate BBM priors with the current prior estimation methods. Second, BBM is currently designed for application to a single population; studies involving multiple groups typically construct priors based on equal representation from each group, but a mixture distribution would arguably better represent the population in this context. \review{Additionally, the application of BBM to studies where participants have differing length or number of sessions remains an open question. Shorter or fewer sessions will generally lead to greater shrinkage of the posterior toward the prior mean, which could introduce spurious differences between participants. For this reason, at this stage we recommend applying BBM to studies where participants have similar session length. We also recommend that users inspect the degree of shrinkage, which is returned by \texttt{fit\_BBM()}, to identify any outliers or major discrepancies between participants. Further research is needed to determine the sensitivity of BBM to scan duration and refine these recommendations.} Third, BBM does not currently incorporate harmonization for multi-site datasets, which is a limitation for consortium datasets like the Alzheimer's Disease Neuroimaging Initiative \citep{ADNI} or the Autism Brain Imaging Data Exchange \citep{di2014autism}. \review{Integrating harmonization into BBM would be beneficial, as harmonization has been shown to improve SNR in resting-state fMRI \citep{chen2022harmonizing, yamashita2019harmonization}.} Finally, while it is possible to incorporate spatial priors into BBM for added accuracy and power \citep{mejia2023template}, the current implementation is computationally intensive, limiting its practical utility.

%\review{Both the potential and the constraints of a Bayesian framework can be observed as the distance between the posterior and the prior distribution, showing how informative the individual data are. A first estimate of shrinkage as a function of posterior values (Figure \ref{fig:app:shrinkage}) indicates that the model follows the prior closer in areas of mean value closer to zero for each network, where individual data is, a priori, not informative about the connectivity of that vertex. As expected, vertices with strong engagement with a given network, do not experience as much shrinkage towards the priors.}

% Future directions/implications for human health
Several areas of future work are planned to address these limitations and extend the capabilities of BBM. These include transfer learning to adapt priors to new populations without large training samples, algorithms for computationally efficient spatial modeling to enhance accuracy and power, inclusion of covariates and nuisance effects such as site and scanner, and mixture priors for heterogeneous populations. All extensions will be implemented in the \texttt{BayesBrainMap} R package as they are developed. We also plan to derive and share topography and connectivity priors from other publicly available resources to facilitate the use of Bayesian brain mapping for personalized functional brain mapping in a wider variety of populations.
\review{Finally, while the scan duration required to produce accurate estimates using BBM (or any other method) depends on numerous factors, precluding a universal recommendation, future work will investigate the sensitivity of BBM to scan duration and assess its ability to reduce the scan time required for accurate mapping of individual functional topography.}

\section{Conclusion}
Bayesian brain mapping (BBM) provides a scalable, flexible, and computationally convenient framework for personalized functional brain network mapping. BBM employs population-derived priors corresponding to established network maps to mitigate noise, thus enhancing accuracy and power while reducing scan duration requirements. Several resources accompanying this manuscript, including the BayesBrainMap R package, pre-estimated HCP-derived priors, and user-friendly demos, lower the barrier to adoption of BBM for personalized functional network mapping. The BBM framework and these openly available resources serve to make precision functional mapping accessible across a wider range of research and clinical contexts.

\section{Acknowledgments}
This work was supported in part by the National Institutes of Health through the National Institute on Aging [Grant No. R01 AG083919 (to A.F.M.)], the National Institute of Mental Health's National Research Service Award [Grant No. F31 MH140552 (to E.R.B.)], and the National Science Foundation Graduate Research Fellowship [Grant No. DGE-2234667 (to E.R.B.)].

% Future work:
% \begin{itemize}
%     \item Transfer learning to adapt priors to different populations (can we use a prior based on healthy young adults to learn a prior for an older population, using a smaller population-specific training set?)
%     \item Dynamic connectivity
%     \item Variational Bayesian techniques for computationally efficient spatial modeling
%     \item Surface and subcortical parcel-constrained spatial modeling with new VB algorithms
%     \item Harmonization of multi-site training data (site effects can inflate between-subject (prior) variance)
%     \item Incorporating covariates in the priors and the model (unmodeled covariate effects can also inflate prior variance, leading to a less informative prior. modeling them could improve subject-level effect estimation)
%     \item Priors for heterogeneous populations (for example, ADNI includes AD, MCI and CU participants)
% \end{itemize}

\newpage

\bibliography{./main.bib}
\bibliographystyle{apalike}

\appendix

\renewcommand\thesection{\Alph{section}}
\renewcommand\thesubsection{\thesection.\arabic{subsection}}
\renewcommand\thetable{S.\arabic{table}}
\renewcommand\thefigure{S.\arabic{figure}}
\setcounter{figure}{0}
\setcounter{table}{0}

\begin{figure}
\centering
\small
\begin{tabular}{C{0.05\textwidth} C{0.30\textwidth} C{0.30\textwidth}}
 &
\makebox[0.30\textwidth][c]{\kern8mm\textbf{Template}} &
\makebox[0.30\textwidth][c]{\kern8mm\textbf{Prior Mean}} \\  
%\hline
\rotatebox{90}{\textbf{Yeo17}} &
\makebox[0.30\textwidth][l]{%
  \kern2mm\includegraphics[width=\linewidth]{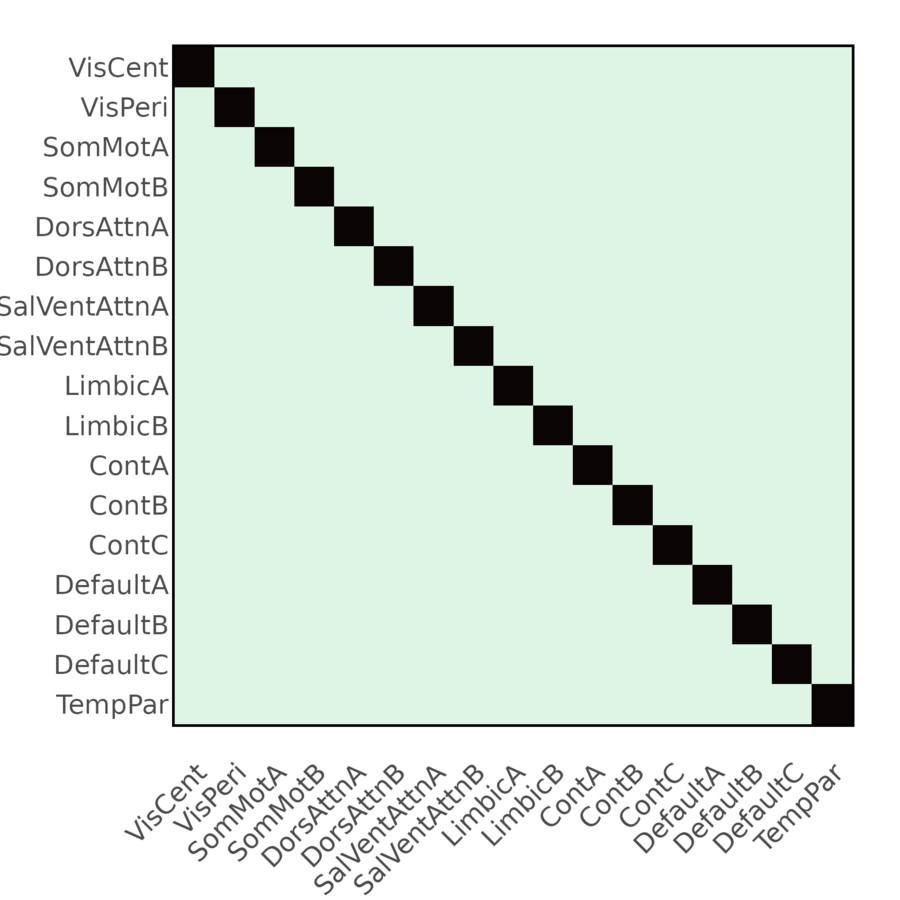}
} &
\makebox[0.30\textwidth][l]{%
  \kern2mm\includegraphics[width=\linewidth]{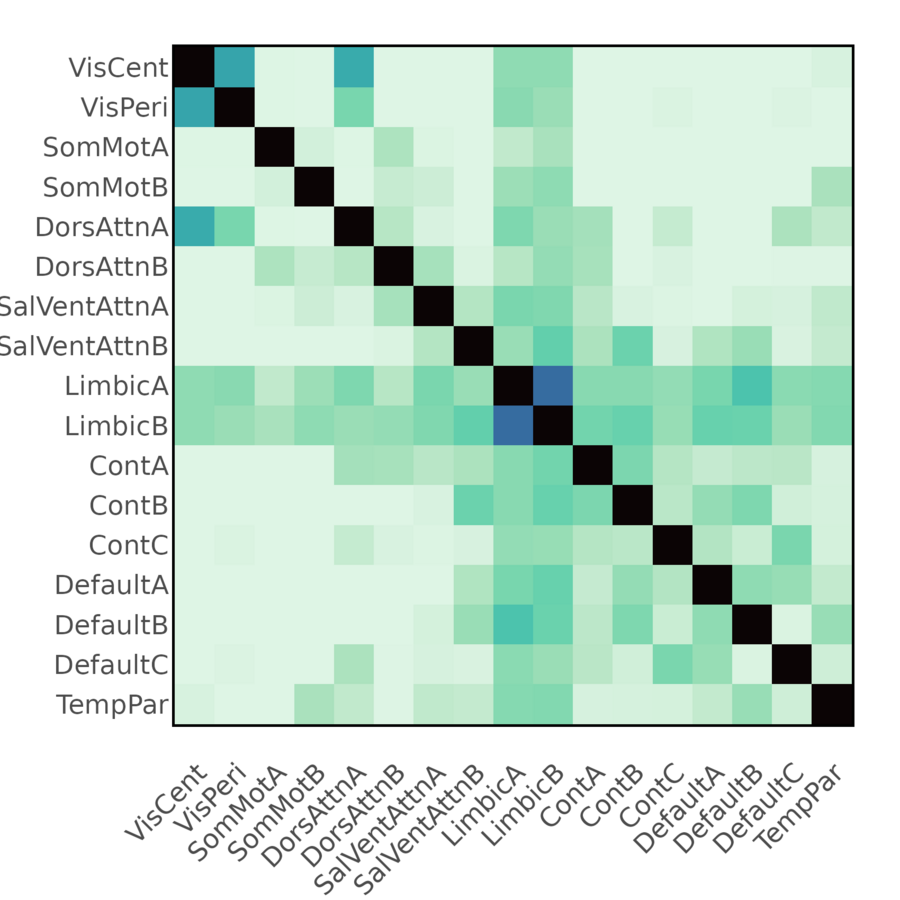}
} \\
\rotatebox{90}{\textbf{MSC}} &
\makebox[0.30\textwidth][l]{%
  \kern2mm\includegraphics[width=\linewidth]{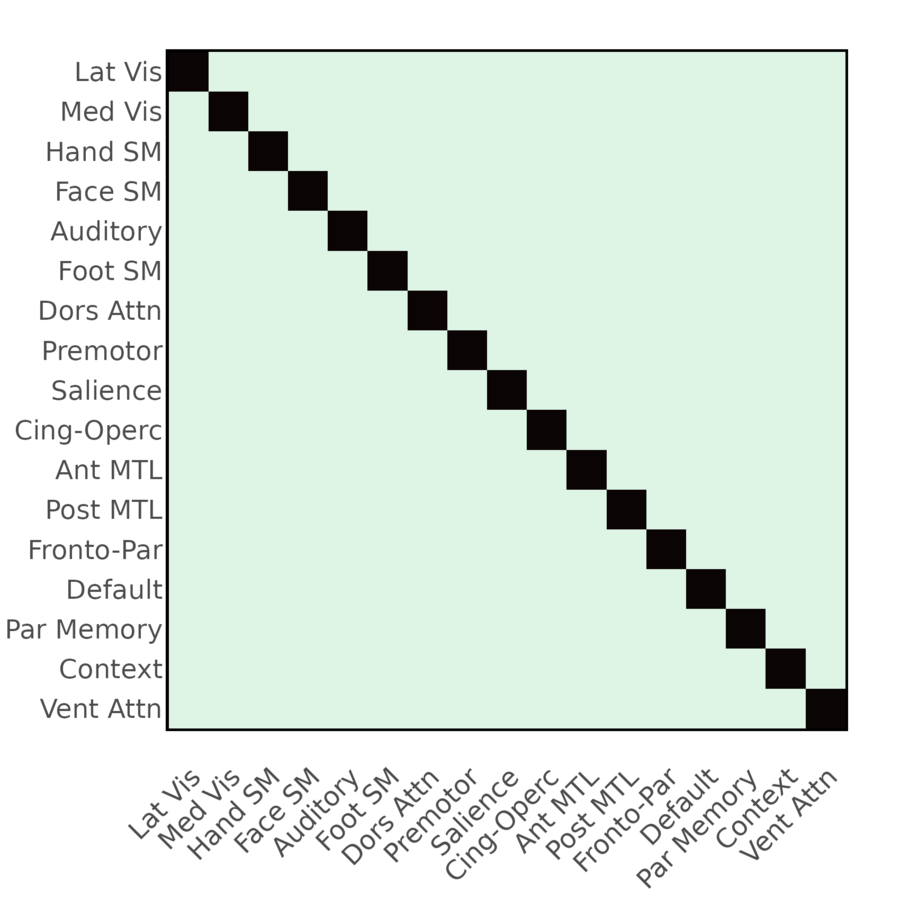}
} &
\makebox[0.30\textwidth][l]{%
  \kern2mm\includegraphics[width=\linewidth]{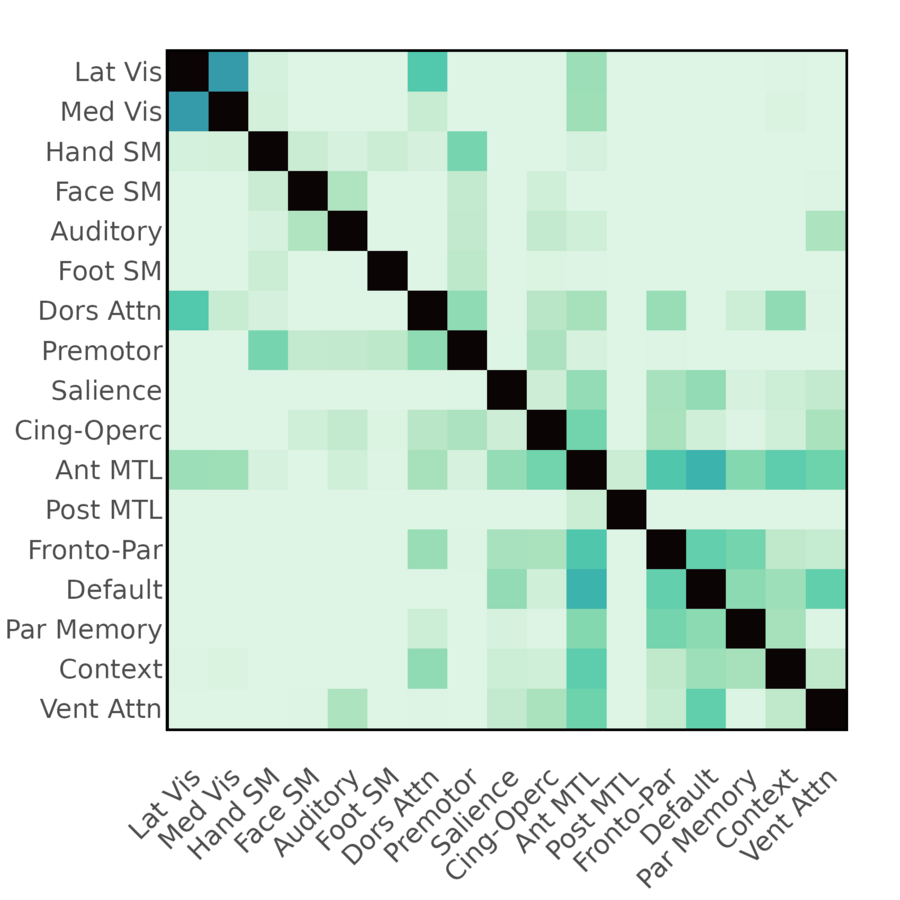}
} \\
\rotatebox{90}{\textbf{GICA15}} &
\makebox[0.30\textwidth][l]{%
  \kern4.5mm\includegraphics[width=0.3\textwidth]{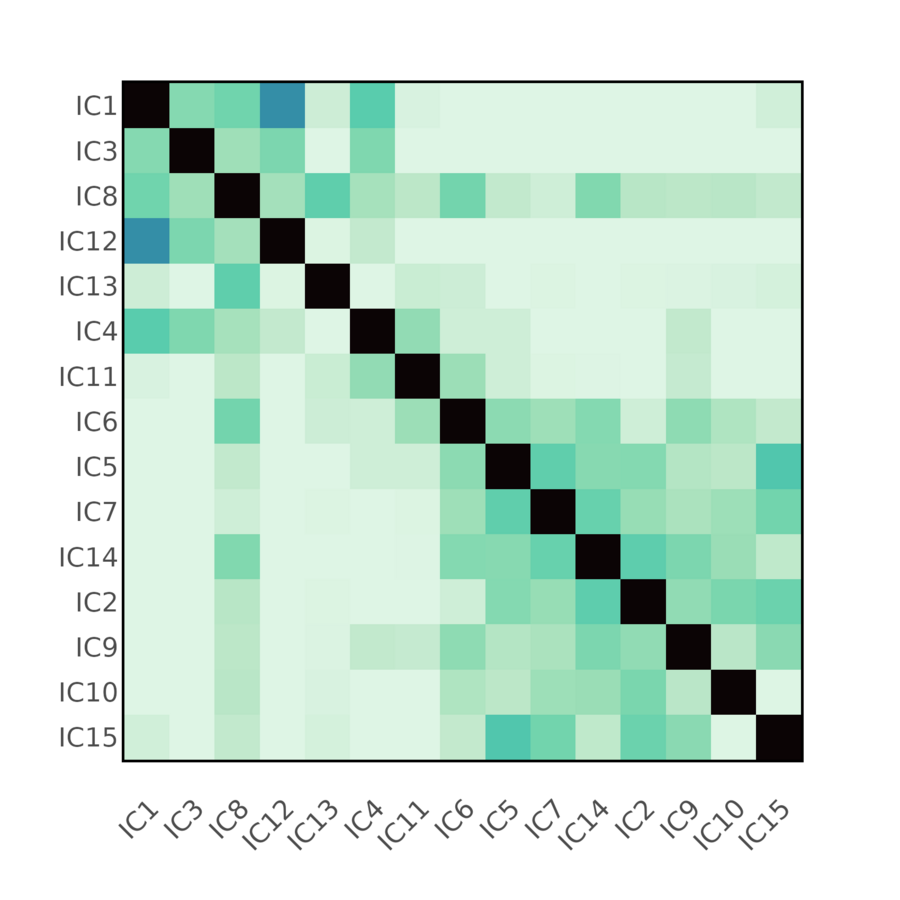}
} &
\makebox[0.30\textwidth][l]{%
  \kern4.5mm\includegraphics[width=0.3\textwidth]{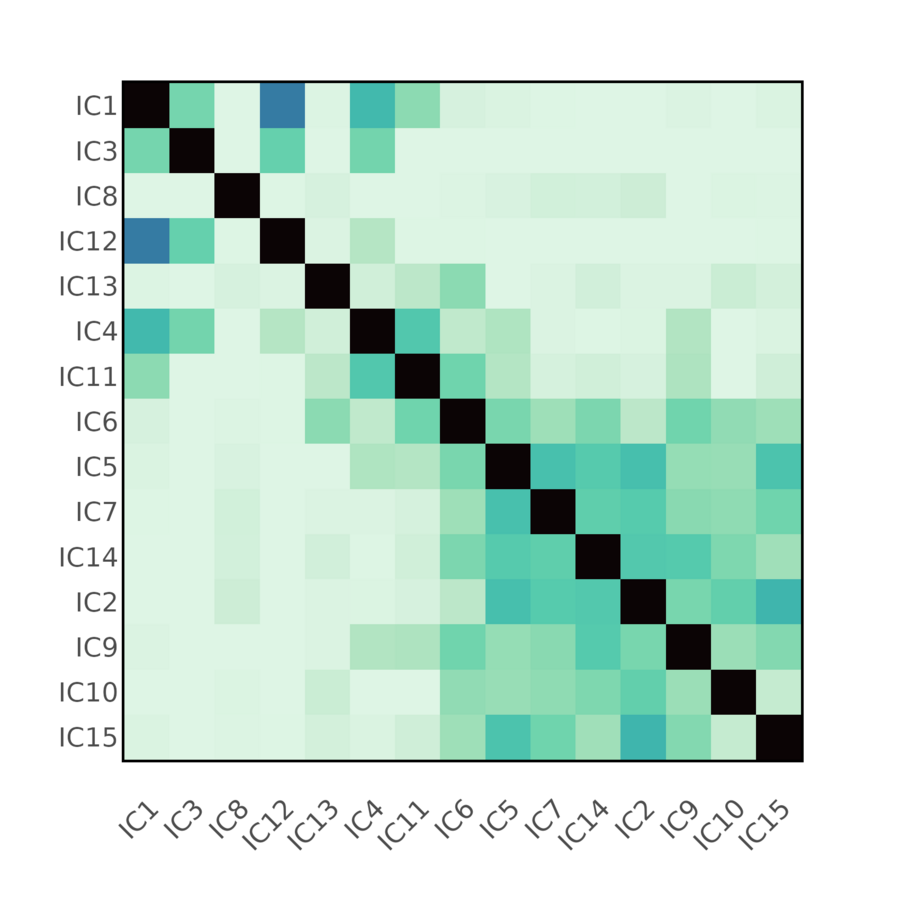}
} \\
\rotatebox{90}{\textbf{PROFUMO}} &
\makebox[0.30\textwidth][l]{%
  \kern2mm\includegraphics[width=0.30\textwidth]{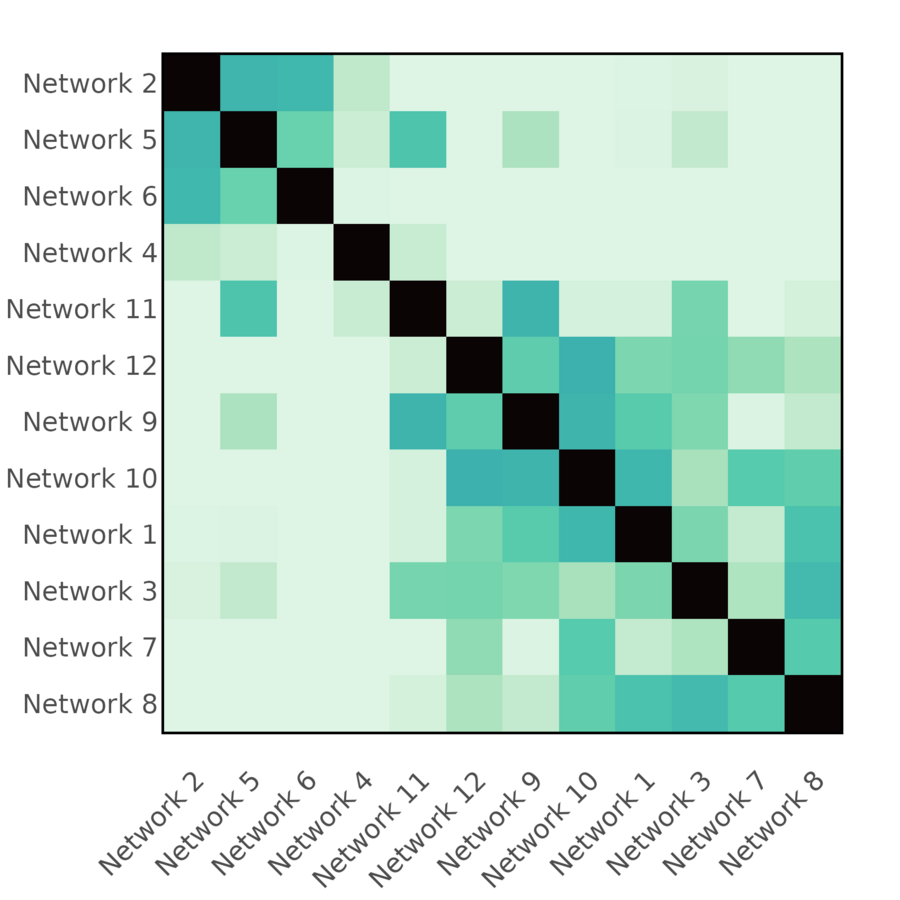}
} &
\makebox[0.30\textwidth][l]{%
  \kern2mm\includegraphics[width=0.30\textwidth]{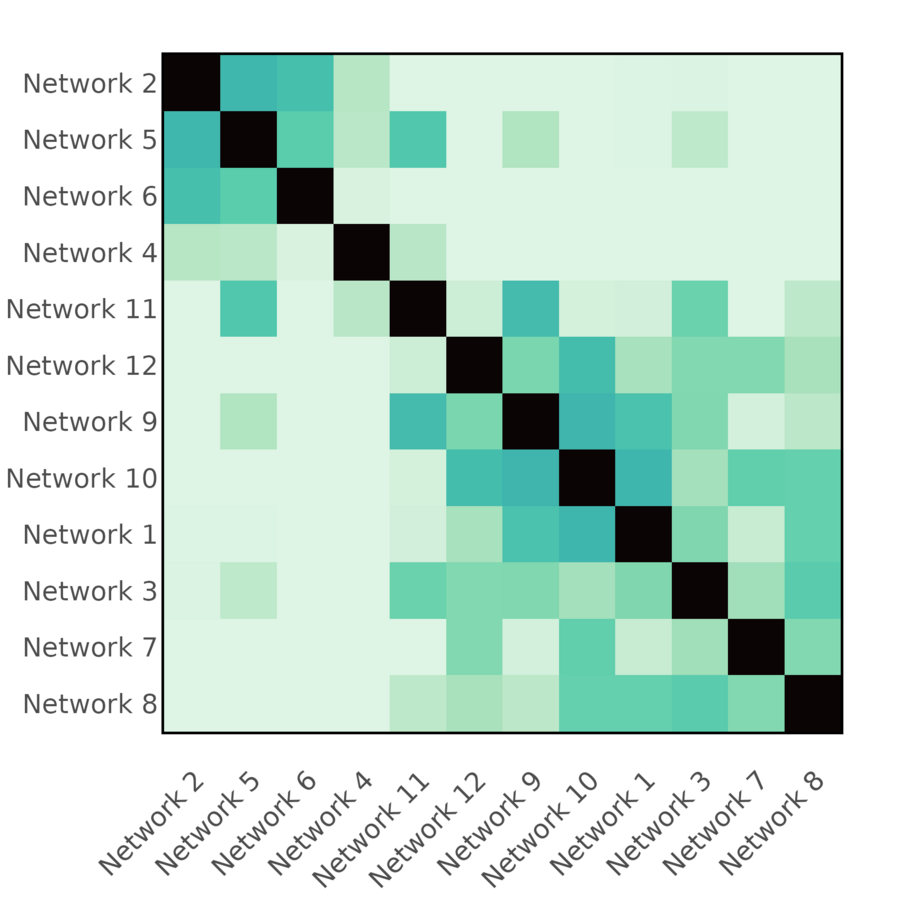}
} \\
\multicolumn{3}{c}{
    \hspace{16mm}\includegraphics[width=0.25\textwidth]{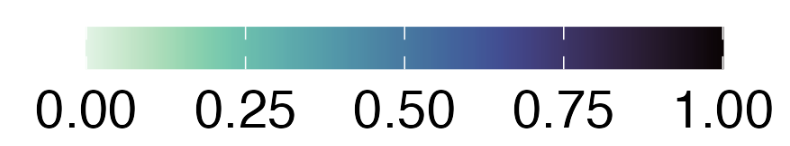}
} \\
\end{tabular}
\caption{Dice overlap matrices comparing each template with its prior mean. Yeo17 and MSC used hard labels; GICA and PROFUMO maps were thresholded at $|z| \geq 2$ before computing Dice. For MSC, GICA and PROFUMO, the networks were matched to the Yeo17 networks based on spatial overlap, so that the network ordering of the matrices is approximately consistent across templates.}
\label{fig:spatial_overlap}
\end{figure}

%TO DO for figure below
% - considering showing the SD also
% DONE - reorder the rows/columns to have a common network ordering 
% - consider removing the Yeo17 based on the way this figure is referred to in the text

\begin{figure}
\centering
\begin{tabular}{cc}
 {Yeo17} & {MSC}\\
\includegraphics[width=55mm, trim = 0 0 0 2cm, clip]{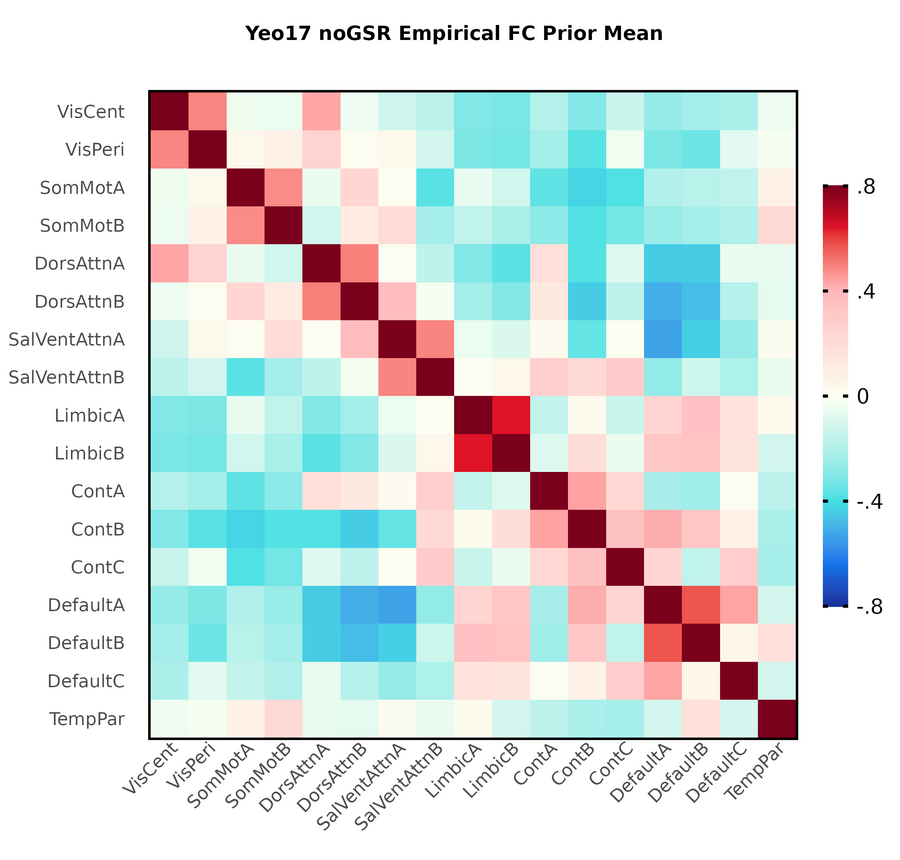} &
\includegraphics[width=55mm, trim = 0 0 0 2cm, clip]{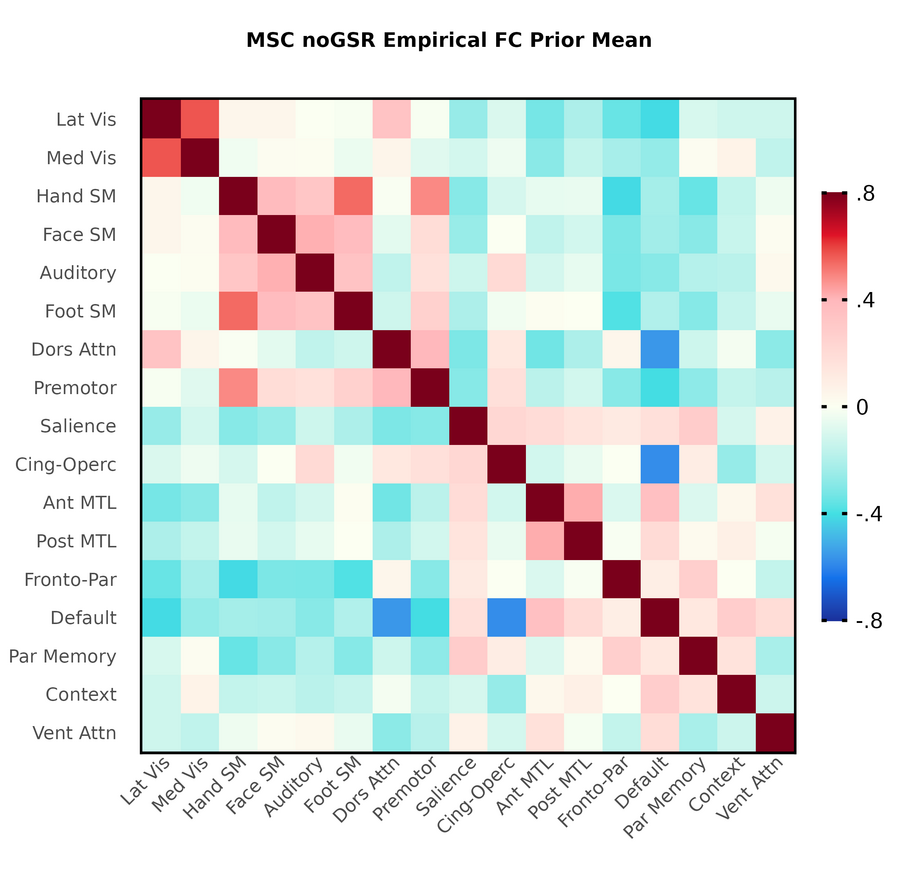} \\
 {GICA15} & {PROFUMO} \\
\begin{picture}(150,140)\put(10,10)
    {\includegraphics[width=5cm, trim = 0 0 0 2cm, clip]{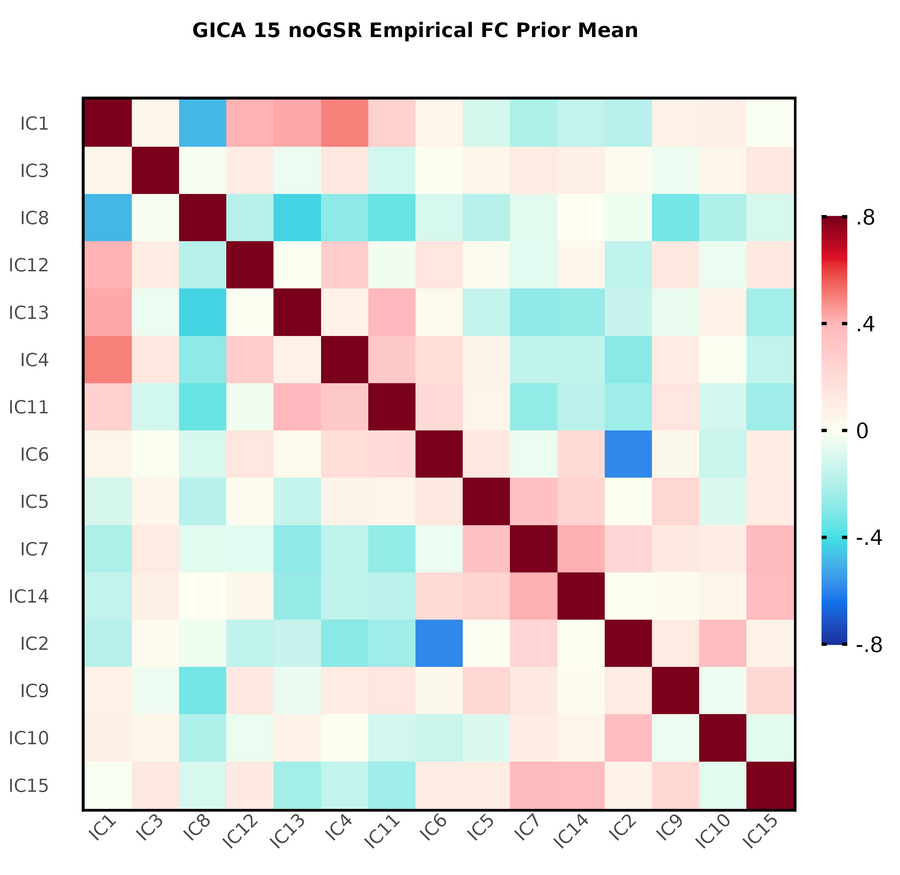}}
\end{picture} &
\includegraphics[width=55mm, trim = 0 0 0 2cm, clip]{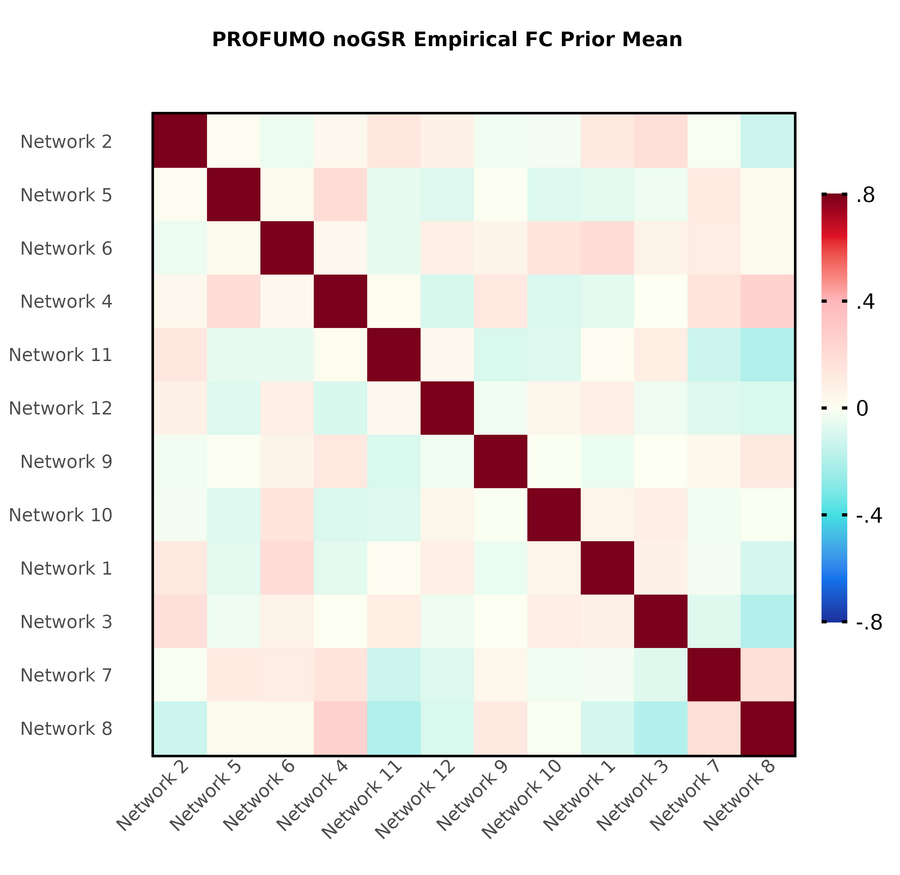} \\
\end{tabular}
\caption{Functional connectivity (FC) priors for various templates. The empirical element-wise population mean of FC across the prior training sample is shown. The empirical population mean and variance are not used directly in the BBM model. Rather, the inverse-Wishart or permuted Cholesky prior is built based on the FC training samples.}
\label{fig:FC_empirical}
\end{figure}

\renewcommand{\RotLabel}[1]{
  \makebox[1.2em][c]{\rotatebox[origin=c]{90}{#1}}
}

\begin{figure}
\centering
\begin{tabular}{@{}c@{\hspace{8pt}}c@{}}
\adjustbox{valign=c}{\RotLabel{Average shrinkage $(\lambda)$}} &
\adjustbox{valign=c}{\includegraphics[width=.9\linewidth]{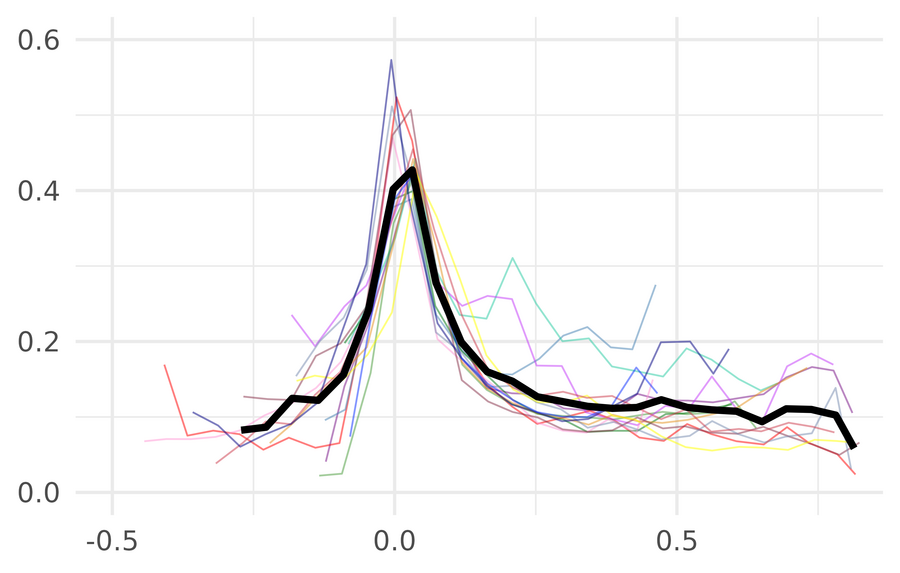}} \\
\ & Average posterior mean
\end{tabular}
\caption{\review{\textbf{Degree of shrinkage by posterior mean value.} For 10 randomly selected HCP subjects, the posterior was calculated using the Yeo 17 priors, and a corresponding MLE was obtained by regressing the network time series against the BOLD data. All vertices were binned according to their posterior mean values, and then plotted against the average shrinkage of those same vertices. Each colored line represents a different network, while the black line represents the average shrinkage across components. The average is only displayed for posterior values represented in at least 5 networks. In background regions for a given network (posterior mean near zero), shrinkage is approximately 40-50\%, but in regions with higher engagement shrinkage is approximately 10\% on average.}}
\label{fig:app:shrinkage}
\end{figure}

\end{document}